# Sedimentary Environment, Diagenesis, Sequence Stratigraphy, and Reservoir Quality of the Ilam Formation in Dezful Embayment and Abadan Plain in South-West Iran

Mahdiyeh Gholizadeh, Mohammad Hossein Adabi[a], Abbas Sadeghi, Mohammadfarid Ghasemi and Maryam Moradi.

[a] Department of Sedimentary Basin and Petroleum, Faculty of Earth Sciences, Shahid Beheshti University Tehran, Iran

## Abstract

The Ilam Formation (Cenomanian–Santonian in age) is considered one of the main rock reservoirs of the Bangestan Group in the southwest of Iran. This formation mostly consists of carbonate rocks. To examine the sedimentary environment, diagenesis, sequence stratigraphy, and reservoir quality of Ilam Formation in Dezful embayment and Abadan Plain, four subsurface sections in wells No. A, B, C, and D, with a range of 106 to 146 meters of thickness, were studied. The lithology of the Ilam Formation in the studied wells is limestone with interbedded shale and argillaceous limestone. Considering the abundance of allochems and various fabrics in these deposits, twelve microfacies and one shale petrofacies of the Ilam Formation were recognized in these four wells. These microfacies were deposited in three facies belts, namely lagoon, shoal, and open marine, in a homoclinal carbonate ramp setting. These deposits have been influenced by meteoric, marine, and burial diageneses. Sequence stratigraphy of the Ilam Formation reveals that the studied wells consist of a third-order sedimentary sequence. The sea-level fluctuations in this area are the same as the global sea-level fluctuations. During the study of the Ilam Formation's reservoir quality based on the results of diagenesis, mainly porosity and permeability data in one of studied wells from the depth of 2850.33 to 2911.13 meters, 6 flow units were identified. Flow unit number 5 has the most potential reservoir quality, and flow unit number 6 has the undesirable flow unit.

## 1. Introduction

The Zagros fold and thrust belt (ZFTB) ,which its approximate length is more than 1500 km ,and its width varies between 100 and 300 km, is located in the NE margin of the Arabian Plate (Stöcklin,1968; Falcon, 1969; Berberian and King, 1981) and is a collisional belt between the Iranian block (belonging to Eurasia) and the Arabian plate, whose convergence started at the beginning of the Late Cretaceous (Berberian and King, 1981). This collision occurred after the total consumption of the Neo-Tethys Ocean (Beydoun et al., 1992; Agard et al., 2005; Mouthereau et al., 2007; Navidtalab et al., 2016; Navidtalab et al., 2019). In the past, Zagros was a part of the Arabian Plate and was located in tropical latitudes, which caused the deposition of carbonate and evaporite sediments in this basin (James and Wynd, 1965). Zagros is one of the most significant basins in the Middle East because of the well-known oil productivity of this region (Afghah, and Farhudi, 2012), and it is a part of the southern margin of Tethys (Stoneley. 1990). From the Lower Cambrian to the Quaternary, a distorted sedimentary package is exposed by the ZFTB. (Stocklin, 1968; Falcon, 1969; Berberian and King, 1981; Sepehr and Cosgrove, 2005). This sedimentary sequence has experienced folding and thrusting during the collision. (Molinaro et al., 2005; Sherkati and Letouzey, 2004; Sherkati et al., 2005).

A sedimentary cycle from Albian to Campanian has been identified composed of Kazhdomi, Sarvak, Surgah, and Ilam formations in Zagros, which are called as Bangestan Group, and name of this group is taken from Bangestan mountain located in the northwest of Behbahan city (James and Wynd, 1965). The most important interval of this group includes neritic carbonates of the Sarvak and Ilam formations and their equivalent units (such as the Mishrif Formation of Iraq). Cretaceous rocks of Zagros do not have the same rock facies and were not deposited in the same sedimentary conditions. Accordingly, the Ilam Formation and its equivalents contain important reservoir intervals in the south and southwest of Iran and throughout the Middle East (Aqrawi et al. 1998; Adabi and Asadi-Mehmandosti 2008; Ghabeishavi et al. 2009). Throughout early Late Cretaceous times, large parts of the Arabian Plate were covered by shallow subtropical seas resulting in the deposition of thick limestone successions ,and these carbonates host a considerable part of the world's total hydrocarbon reserves (Taghavi et al., 2006; Beiranvand et al., 2007). Despite their outstanding economic importance, the stratigraphic assignment of the Cretaceous neritic carbonates is notoriously difficult due to the absence of typical open-marine index fossils (Omidvar et al., 2014).

Carbonate rocks are non-clastic sedimentary rocks that contain more than 50% of carbonate minerals. Investigating the petrography of carbonate rocks, which includes texture and structure, and identifying the main and secondary constituents of sedimentary facies are important in investigating microfacies and changes in the sedimentary environment. Environmental factors such as depth, temperature, salinity, biological substrate and disturbance, dispersion and spread of organisms are among the controlling factors of carbonated environments (Tucker, 2001).

Carbonate rocks are sensitive to diagenesis changes and the replacement of crystals occurs over time by temperature and pressure. In thin sections, this replacement of crystals can be seen with changes in size, crystal shape, colors, and the creation of impurities compared to the original grains (Wilson, 2012), but the carbonate facies sequences are the result of environmental changes over time (Tucker and Wright, 1990). As a result, by identifying the microfacies of carbonate rocks, the characteristics and conditions of their formation can be interpreted and a schematic model for the paleo-sedimentary environment of these sediments can be presented.

The Cretaceous geological system is one of the most important systems in the history of geology, especially in the geology of Iran in the Zagros fold region. This is due to the fact that various sedimentary conditions led to the desirable setting for the formation of oil traps. Facies variations of the Ilam Formation in horizontal and vertical directions as well as its diagenetic changes, led to different reservoir qualities in the formation in various areas of Zagros. In this regard, the present study aims to investigate the sedimentary environment, diagenetic processes, sequence stratigraphy, and reservoir characteristics of this formation in subsurface sections (wells No. A, B, C, D) located in the southwest of Iran.

## 2. Geological setting

The study area is located in the Dezful Embayment and Abadan Plain southwest of Iran (Fig. 1), which is part of the Zagros fold-and-thrust belt (Alavi, 2004). The Dezful Embayment is located southwest of the Zagros thrust (Alavi, 2007). Despite its relatively small area, the Dezful Embayment produces a large portion of Iran's oil reserves (Bordenave and Huc, 1995) because the Gachsaran Formation has covered Asmari limestone as a cap rock in the oil system (Berberian, 1995). The Dezful Embayment is separated from other zones by the Mountain Front Fault (MFF), Balarud Fault Zone (BFZ), Izeh Fault Zone (IFZ), Kazerun Fault Zone (KFZ), and Zagros Front Fault (ZFF), Zagros Main Reverse Fault (ZMRF), High Zagros Fault (HZF), Zagros Deformation Front (ZDF) (Fig. 2).

The Abadan Plain is located in southwest Iran and is surrounded by the Dezful Embayment, the Persian Gulf, and the Iran-Iraq boundary (Fig. 2). It is a part of the Mesopotamian Basin, and its structural characteristics differ from the Dezful Embayment (Zeinalzadeh et al., 2015). Mesopotamian Basin is considered one of the richest petroleum systems in the world (Sadooni and Aqrawi, 2000). The Mesopotamian Basin is a vital hydrocarbon province in Iraq and contains several well-known oil fields. Also, Abadan Plain has some good oil fields that are considered for future development and production. A thick Mesozoic succession and ranges dominate the stratigraphic section in the Abadan Plain from Jurassic to Cretaceous. The Cretaceous rocks in the basin are considered gas and oil reservoir rocks (Zeinalzadeh et al. 2015) and comprise fractured and vuggy carbonates as well as clastic rocks. The Abadan Plain has been relatively less influenced by the tectonic compression resulting from the closure of the Neo-Tethys Ocean (Abdollahie-Fard et al., 2006; Aqrawi and Badics, 2015). Consequently, it is structurally characterized by gentle folding compared to more eastward sub-basins in SW Iran (Aqrawi and Badics, 2015). In this plain, similar to the other parts of the Mesopotamian Basin, due to the cover of recent alluvial deposits and the absence of outcrops, the geological information is mainly limited to the subsurface, including drilling boreholes and geophysical data (Sissakian, 2013).

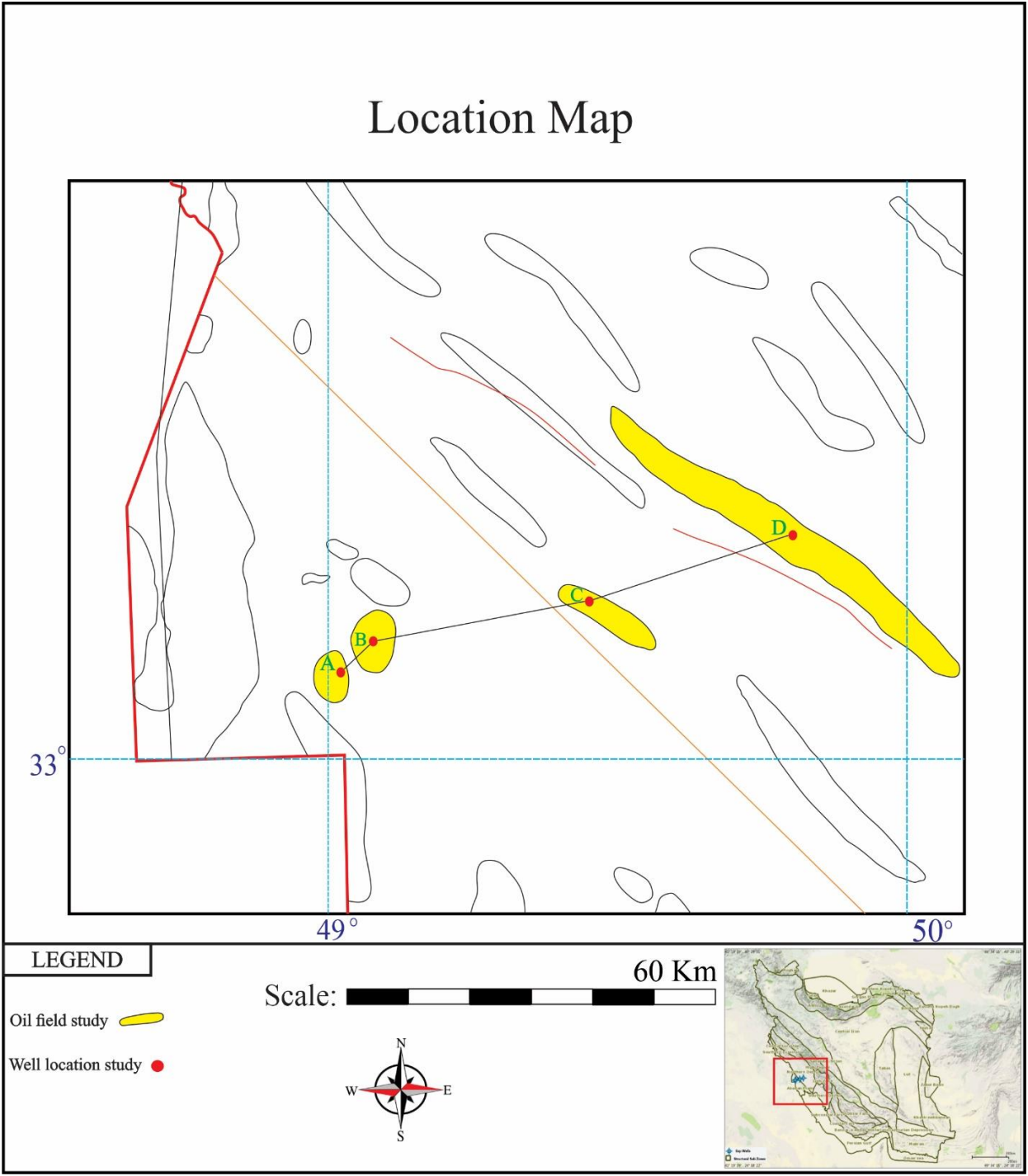

Fig. 1. The location of the studied oil fields.

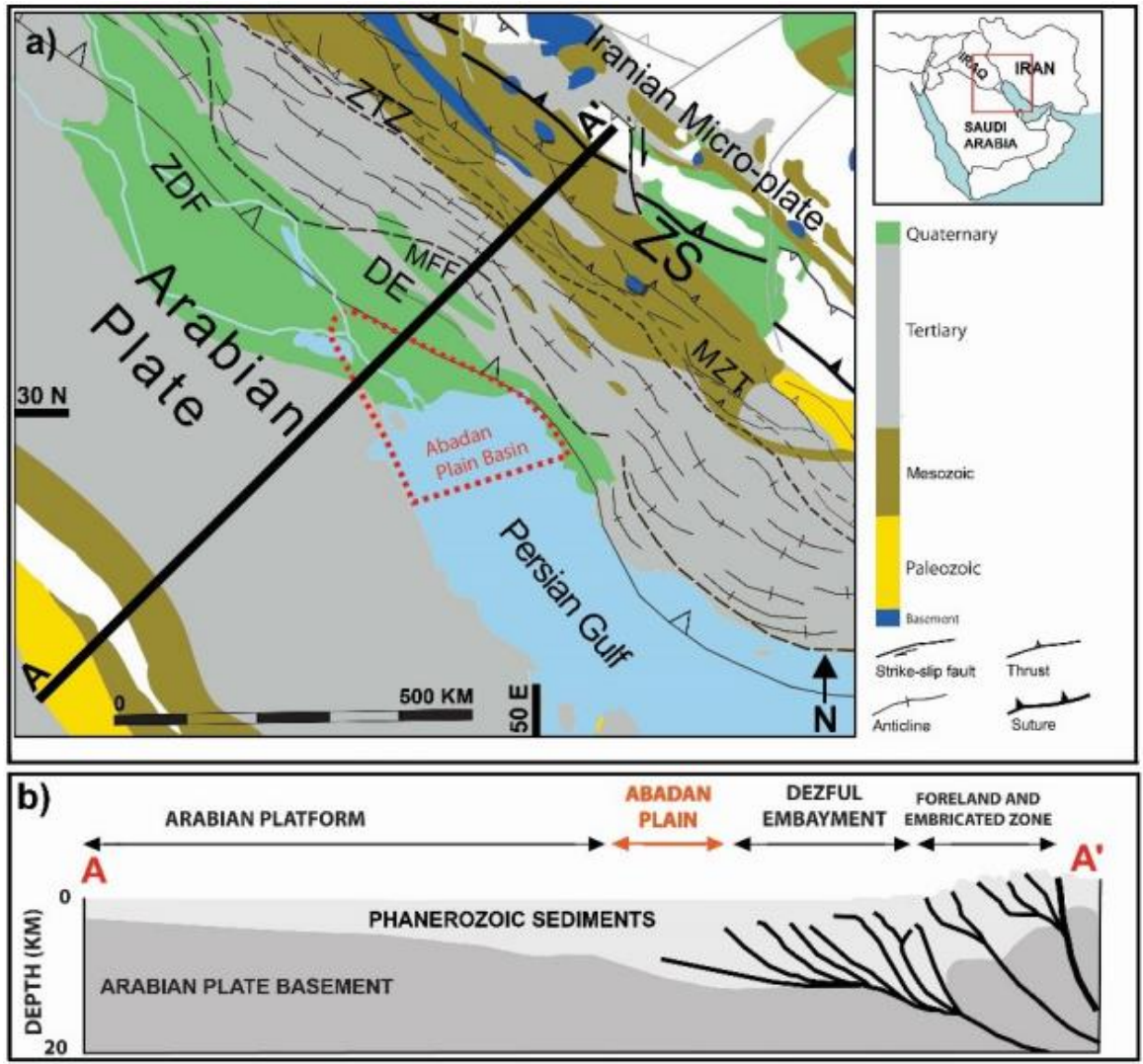

Fig. 2. (a) geological map of the Arabian Plate and Zagros Fold-Thrust-Belt. Tectonic features are from reference (Alavi, 2007). Abbreviations: DE: Dezful Embayment (the region between the Abadan Plain Basin and MFF), MFF: Mountain Front Flexure, MZT: Main Zagros Thrust, ZTZ: Zagros Thrust Zone, ZDF: Zagros Deformation Front, ZS: Zagros structure. The Zagros front fault bounds Abadan Plain Basin to the north and northeast. (b) cross-section across the Arabian Shield, Arabian Platform, and the ZFTB along line A–A' modified from reference (Abdollahie-Fard et al., 2006), and all final changes are from reference (Atashbari et al., 2018).

## 3. Materials and methods

This study is based on the subsurface data (Fig. 1) and basic information, including thin sections of cores, cuttings, and sonic and gamma-ray logs in the four studied wells. The petrographic analysis includes examining and identifying carbonate and non-carbonate components, identifying diagenetic processes, and recognizing different microfacies. In this study, 388 thin sections from the Ilam Formation have been examined.

Furthermore, Dunham's (1962) classification and Burchette and Wright's (1992) model were used for facies analysis and to create a sedimentary environment model, respectively. In addition, the identification of dolomite types has been carried out, according to Adabi (2009). Classification and identification of pore spaces have been done according to Choquette and Pray (1970) and Lucia (1983). Meanwhile, Hunt and Tucker's (1992) model has been applied to determine depositional sequences and sequence boundaries. The determination of reservoir quality and flow units of the Ilam Formation in the one of subsurface section is based on the data from cores, logs, and thin sections using the Kozeny-Carman equation (Carman, 1937; Kozeny, 1972), Amaefule et al. method (Amaefule et al., 1993), and Gaunter method based on the Winland and Lorenz equation (Gunter et al., 1997).

## 4. Results and discussion

### 4.1. Facies analysis

Identification of microfacies and facies belts is one of the main parts of reservoir geology study in comprehensive reservoir studies (Lucia, 2007; Ahr, 2008; Moore and Wade, 2013). The microfacies description of the Ilam Formation has been carefully presented in some past studies in the Zagros basin (Adabi and Asadi-Mehmandosti, 2008; Gabeishavi et al., 2009; Mehrabi et al., 2013; Rahimpour-Bonab et al., 2013; Khodaei, 2021). Based on the petrographic studies, four subsurface sections of the Ilam Formation sequence, located in southwest Iran, were examined from twelve carbonate microfacies and one shale petrofaceis. The recognized of 388 thin sections from the studied subsurface sections with different thicknesses led to the identification of carbonate microfacies in three facies belts, which are presented from deep to shallow sub-environments:

### 4.1.1. Open marine facies belt

Most of the skeletal components of open marine facies belts are sensitive to the salinity of seawater (Flugel, 2016). In this part of the sedimentary environment of the Ilam Formation, mudstone, wackestone, and packstone microfacies consist mainly of bioclast planktonic foraminifera. The microfacies of this facies belt have a micritic matrix. Overall, the studied microfacies of this facies belt including:

### MF1: Bioclast oligostognid planktonic foraminifera wackestone to packstone

This microfacies is found in the lower part of the Ilam Formation adjacent to the Sarvak Formation. Allochems in this facies, including ostracods, bivalves, gastropods, bryozoan, textularide, and echinoid debris, are found with less than 10 percent in abundance. Planktonic foraminifera includes heterohelix, small rotaliids, hedbergella, and oligostenginid group with 15 percent in abundance. This facies' most significant diagenetic processes are dolomitization, particularly around stylolites and minor iron oxide. Interparticle porosity caused by foraminifera dissolution and fracture can be observed to some extent. This microfacies is observed in well No. A (Fig. 3 A).

**Interpretation:** Mud matrix, the existence of planktonic foraminifera along with echinoid debris, and presence of oligostenginid suggested that this microfacies belongs to the relatively deep sedimentary basin below the storm wave base (SWB). This microfacies is similar to the standard Ramp microfacies (RMF) 3 (Flugel, 2016) and belongs to the open marine (outer ramp setting). The same facies are generally described from outer platform areas of carbonate shelfs and ramps (Bauer et al., 2002; Ghabeishavi et al., 2009; Mehrabi et al., 2013).

## MF2: Planktonic foraminifera wackestone

This microfacies is found at the lower parts of the Ilam Formation in well No. A. This microfacies with a micritic matrix contains planktonic foraminifera, including heterohelix, hedbergella, globigerinelloides, and oligostenginid fossils with a 10 to 15 percent in abundance. In this microfacies, diagenetic processes, including cementation in foraminifera chambers, formation of calcite cement in fractures, pyritization, and chemical compaction, can be observed. This microfacies is observed in well No. A (Fig. 3 B).

**Interpretation:** The evidence of bioturbation and traces of clastic grains, the existence of carbonate mud in large quantity, as well as the existence of planktonic foraminifera indicate the sediment of this facies formed below the storm wave base (SWB) with a low energy environment. This microfacies is the same as the standard microfacies RMF 3 introduced by Flugel (2016), and it has been attributed to the outer ramp setting.

## MF3: Planktonic foraminifera mudstone to wackestone

The main components of this microfacies including planktonic foraminifera and oligostenginid fossils with an abundance of about 10 percent. The abundance of planktic foraminifera and oligostegina in a mud-supported texture indicates deposition in a low-energy, deep marine setting (Bauer et al. 2002; Schulze et al. 2005). Other components, such as echinoderms and bivalve debris, can be observed in this microfacies. Micritization, dolomitization, pyritization (inside foraminifera chambers), bioturbation, chemical compaction, and cementation (pore filling, drusy, blocky types) are among the diagenetic processes that can be observed in this microfacies. Porosities such as fracture, interparticle, and vuggy porosities are present in the facies. This microfacies is observed in wells No. A, B, C, and D (Fig. 3 C).

**Interpretation:** The presence of dominant mud texture, planktonic foraminifera, and very low abundance of benthic foraminifera indicate this microfacies is belong to the RMF 2 and deeper part of the basin below storm wave base (SWB) (Flugel, 2016), which belongs to the outer ramp settings.

## MF4: Argillaceous mudstone/ limestone

This microfacies has a mud matrix. Planktonic foraminifera in this microfacies have a low abundance. Among the diagenetic processes in this microfacies, sporadic dolomitization can be observed. This microfacies is present in wells No. A and B (Fig. 3 D).

**Interpretation:** Mud matrix and presence of planktonic foraminifera reveal that this microfacies has been formed in the deep part of the basin with low energy below the storm wave base (SWB). This microfacies is similar to the standard microfacies RMF 2 introduced by Flugel (2016), and it has been attributed to the outer ramp.

**MF5: Planktonic foraminifera bioclast wackestone to packstone**

The major components in this microfacies are: bioclasts (Bivalves, gastropods, echinoids, and ostracods) with a 25 percent abundance. Planktonic foraminifera and oligostenginid fossils are other skeletal grains with 10 percent in abundance, which are present in a micritic matrix. The diagenetic processes in this microfacies include: dissolution, cementation (drusy, equant, microgranular, and syntaxial), pyritization, chemical compaction, physical compaction, bioturbation, and micritization. This microfacies, due to severe dissolution, contains a high percentage of porosity (such as fracture, channel, vuggy, moldic, and matrix porosities). This microfacies is observed in well No. A (Fig. 3 E).

**Interpretation:** The high abundance of skeletal debris in various sizes and carbonate matrix indicates a low-energy environment. The presence of planktonic foraminifera shows that this microfacies has been formed close to the mid-ramp settings. This microfacies is similar to the standard microfacies of RMF 8 introduced by Flugel (2016).

**4.1.2. Shoal facies belt**

Shoal facies belt during sedimentation has continuously been under the influence of waves and currents. Ooid facies formed in warm waters with a depth of fewer than 5 meters, having a salinity of a little higher than the normal level, and with relatively severe turbulence suggest a high energy environment of the shoal facies belt (Tucker, 2001; Flugel, 2016). Allochems of this facies belt include ooid, intraclast, and pelloid in packstone to grainstone textures. In general, the percentage of non-skeletal allochems in some facies is more than 50 percent. Ooids in this facies have undergone deformation due to physical compaction and, in some cases, dolomite replacement and oxide iron formation.

**MF6: Ooid intraclastic bioclast packstone to grainstone**

This microfacies contains about 15 percent intraclast and a small amount of pelloid, coated grains (cortoid), and ooids. Various types of deformed and micritizied ooids have been observed in this microfacies. Echinoid debris and other skeletal debris are present. The diagenetic processes include bioturbation, pyritization, micritization (probably due to algal activity), and cementation. Pore-filling cement types are in the form of drusy and blocky cements. This microfacies is observed in wells No. A, B, and C (Fig. 3 F).

**Interpretation:** This microfacies mainly consists of carbonate mud and some sparry calcite cement. The large sizes of the depositional components, rounded allochems, and grainstone to packstone texture suggest relatively high-energy depositional settings in the leeward part of the shoal environment. This microfacies is similar to the standard microfacies of the RMF 26 ramp model introduced by Flugel (2016).

**MF7: Intraclastic bioclast grainstone**

This microfacies contains about 25 percent intraclast and a small amount of pelloid and coated grains (cortoid). Echinoid debris is of the most important bioclasts that are found in this microfacies, which have been micritizied to some extent. The diagenetic processes in this microfacies include compression, fracture, micritization, and cementation. Pore-filling cement types are in the form of drusy and blocky cements. This microfacies is observed in well No. A (Fig. 3 G).

**Interpretation:** The matrix of this microfacies is made up of sparry calcite cement and a small amount of carbonate mud. The large sizes of the depositional components, the size of rounded allochems, and grainstone texture suggest a high-energy depositional environment.

This microfacies belongs to the central part of the shoal, indicating a high-energy environment and subjected to waves and currents; thus, it is considered shoal facies (Flugel, 2016). This microfacies is similar to the standard microfacies RMF 27 introduced by Flugel (2016) and belongs to the high energy part of the shoal and adjacent to the middle ramp.

**MF8: Ooid grainstone**

This microfacies, contains 30 percent dark ooid with a radial tangential fabric with good sorting in a sparry calcite matrix and sometimes in a carbonate mud matrix. These non-skeletal allochems are rich in iron oxide. In addition, small amounts of coated grains, pelloids, and intrclasts can be found in this microfacies. Ooids are seen in different shapes, including surface, radial, tangential, and micritizied ooids. The existence of ooids with proper sorting and radial fabric shows original calcite mineralogy (Tucker, 2001). Skeletal debris such as foraminifera, echinoid debris, rotalia, gastropods, and bivalve form the core of the majority of ooids. Diagenetic processes in this microfacies are compaction (which resulted in the deformation of some of the ooids), cementation (drusy, pore-filling, blocky, isopachous, needle-shaped, and bladed types), selective dolomitization in some of the ooids, and micritization. Ferrugenation in ooids is another diagenetic process. This microfacies is observed in wells No. A, B, C, and D (Fig. 3 H).

**Interpretation:** The presence of ooids in a grainstone texture and the location of this microfacies at the top of Fair weather wave base FWWB and in the high-energy environment suggest a shoal sub-environment (Flugel, 2016). This microfacies is the same as the standard microfacies RMF 29 introduced by Flugel (2016) and belongs to the inner ramp's central part of the high-energy shoal environment (Tucker and Wright, 1990; Insalaco et al., 2006).

 Well- to very well sorting and high roundness of grains indicate that this facies belongs to the high energy shoal environment (Bauer et al. 2002; Schulze et al. 2005; Blomeier et al. 2009; Jamalian et al. 2011; Mehrabi et al. 2013).

**MF9: Bioclast pelloidal packstone to grainstone**

Pelloids with more than 40 percent in abundance, form the main carbonate components of this microfacies. These non-skeletal allochems lack internal structure, and some of them are micritizied ooids. Ooids in this microfacies have good roundness and sorting. The compaction of pelloids in different parts of this microfacies is variable, and spaces are filled with sparry calcite cement. Benthic foraminifera, rotalia, and echinoids are some of the skeletal allochems seen in this microfacies. Cementation and micritization are essential diagenetic processes. This microfacies is observed in wells No. A, B, and D (Fig. 3 I).

**Interpretation:** The large amount of pelloids with grain-supported fabric suggests a lower energy environment; however, the existence of cement among intraclasts and pelloids with a good sorting could be a sign of a relatively high energy shoal environment in the direction of seaward shoal (Tucker and Wright, 1990). Therefore, this microfacies has been deposited in an area between the lagoon and shoal. Based on the studies, it can be said that this microfacies is the same as standard microfacies RMF 26 introduced by Flugel (2016) and belongs to the lower energy seaward shoal environment in the near mid ramp settings.

### 4.1.3. Lagoon facies belt

The lagoon facies belt is located in the back part of the shoal, which is separated by ooid or bioclast shoal (Flugel, 2016). The energy of this area is low, and sediments have been deposited in a quiet area, therefore, the matrix of the sediments is made up of carbonate mud. Microfacies in this facies belt include MF10 to MF12, and shale petrofacies contain benthic foraminifera such as miliolid with a porcelaneous shell and rotalia. Non-skeletal allochems such as pelloids in this part of the Ilam Formation can be found in a higher amount. Echinoid debris, bryozoan, bivalve debris, and red and green algae are among the most important skeletal allochems present in the sediments of the lagoonal sub-environment.

### MF10: Mudstone

Skeletal allochems in this microfacies have a small abundance and include benthic foraminifera with less than 5 percent. Solution seams and stylolites along dead oil with a dark color are caused by compaction. This microfacies is observed in wells No. C and D (Fig. 4 A).

**Interpretation:** Lack of any association with evaporite facies and the presence of mud in large volume in this microfacies indicates that it has been deposited in a quiet environment and away from wave base shore. This microfacies is similar to the microfacies RMF 19 introduced by Flugel (2016) and belongs to the lagoonal environment.

### MF11: Pelloid echinoid bioclastic wackestone to packstone

Pelloids with about 15 percent have well sorting, and echinoids with about 10 percent abundance are observed in this microfacies . Other skeletal allochems include rotalia and benthic foraminifera, such as textularide, which form a small percentage (about 10 percent) of this microfacies. Based on the petrographic studies, the most important diagenetic processes include micritization, pyritization, dolomitization, and limited cementation. This microfacies is observed in well No. A (Fig. 4 B).

**Interpretation:** The lack of a large amount of biological debris and micrite suggests a low-energy environment. A high abundance of pelloid and mud supported texture indicates its deposition in a confined environment such as a low-energy lagoon. This microfacies is the same as standard microfacies RMF 20 introduced by Flugel (2016) and belongs to the inner ramp.

## MF12: Benthic foraminifera bioclast wackestone

The most essential components of this microfacies are skeletal allochems, especially benthic foraminifera, including miliolid, textularide, and nezazzata, with an abundance of about 10 percent which are present in a carbonate mud matrix. Pelloids in this microfacies have an abundance of less than 5 percent. Dolomitization, micritization, pyritization, and limited cementation are diagenetic processes. This microfacies is observed in wells No. A, B, C, and D (Fig. 4 C).

**Interpretation:** This microfacies with a mud supported matrix contains benthic foraminifera such as miliolid with a porcelaneous coating which belongs to low energy restricted lagoonal depositional environment. This microfacies is similar to standard microfacies RMF 20 introduced by Flugel (2016) and belongs to the lagoonal environment in the inner ramp settings. Similar facies are commonly attributed to the shallow lagoonal to tidal settings in carbonate platforms (Harris, 2009)

## PF1: Shale petrofacies

In the sequences located at the beginning of the studied wells, shallow depth shale depositions in the form of interlayers in the limestone sequences of the Ilam Formation can be found. Sonic and gamma logs reveal a very high increase in the studied shale sequences. The increase in the gamma log can be due to the presence of shale layers or the presence of iron oxides in this petrofacies. Based on the petrographic studies, thin sections of limestone around these shale units belongs to the lagoonal shallow depth depositional environment, therefore, these shale sequences most probably belong to the lagoonal environment (Fig. 4 D).

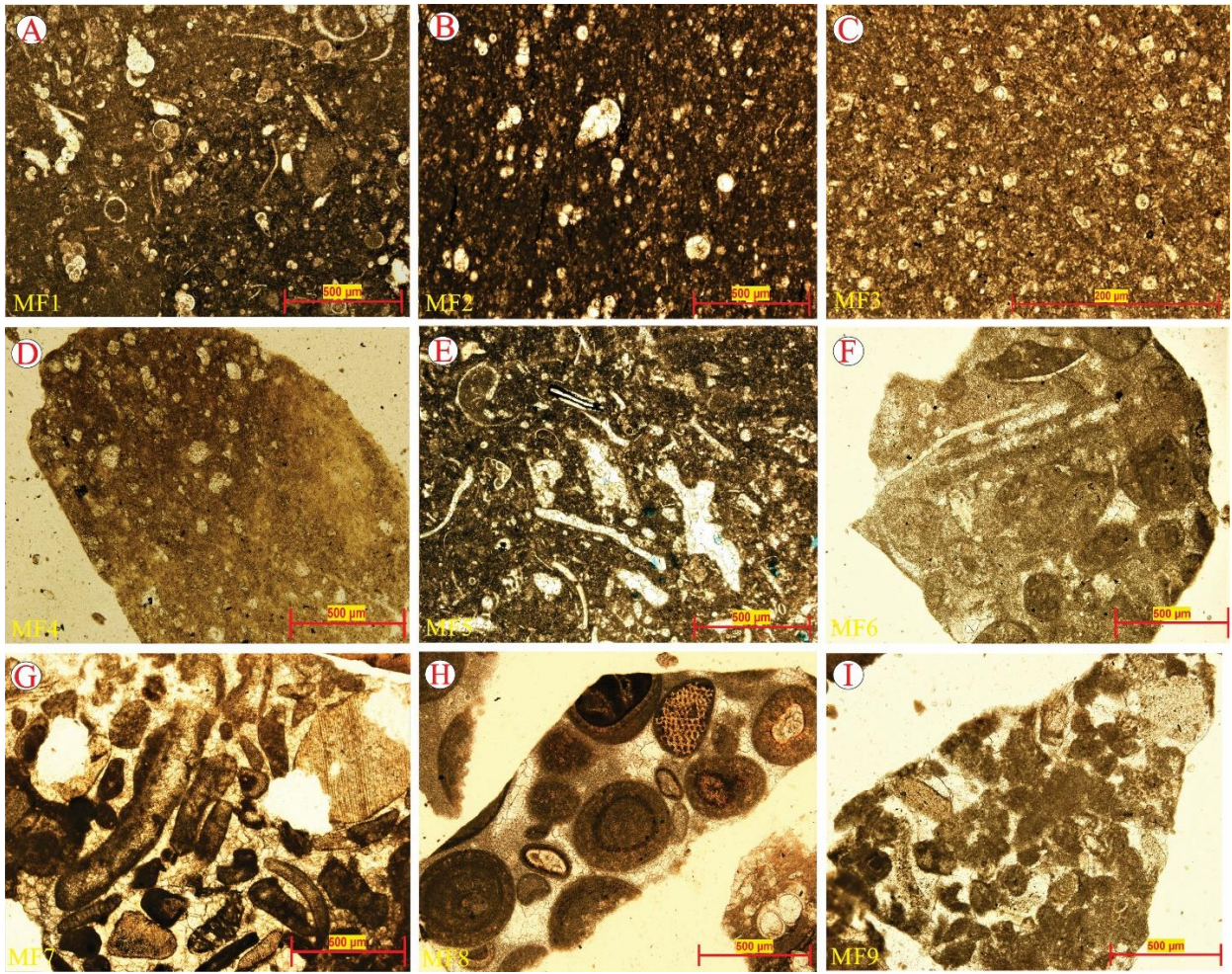

Fig. 3. (A). Bioclast oligostognid planktonic foraminifera wackestone to packstone (well A, PPL). (B) Planktonic foraminifera wackestone (well A, PPL). (C) Planktonic foraminifera mudstone to wackestone (well A, PPL). Argillaceous mudstone/limestone (well B, PPL). (E) Planktonic foraminifera bioclast wackestone to packstone (well A, PPL). (F) Ooid intralclast bioclast packstone to grainstone (well D, PPL). (G) Intraclast bioclast grainstone (well A, PPL). (H) Ooid grainstone (well D, PPL). (I) Bioclast pelloid packstone to grainstone (well D, PPL).

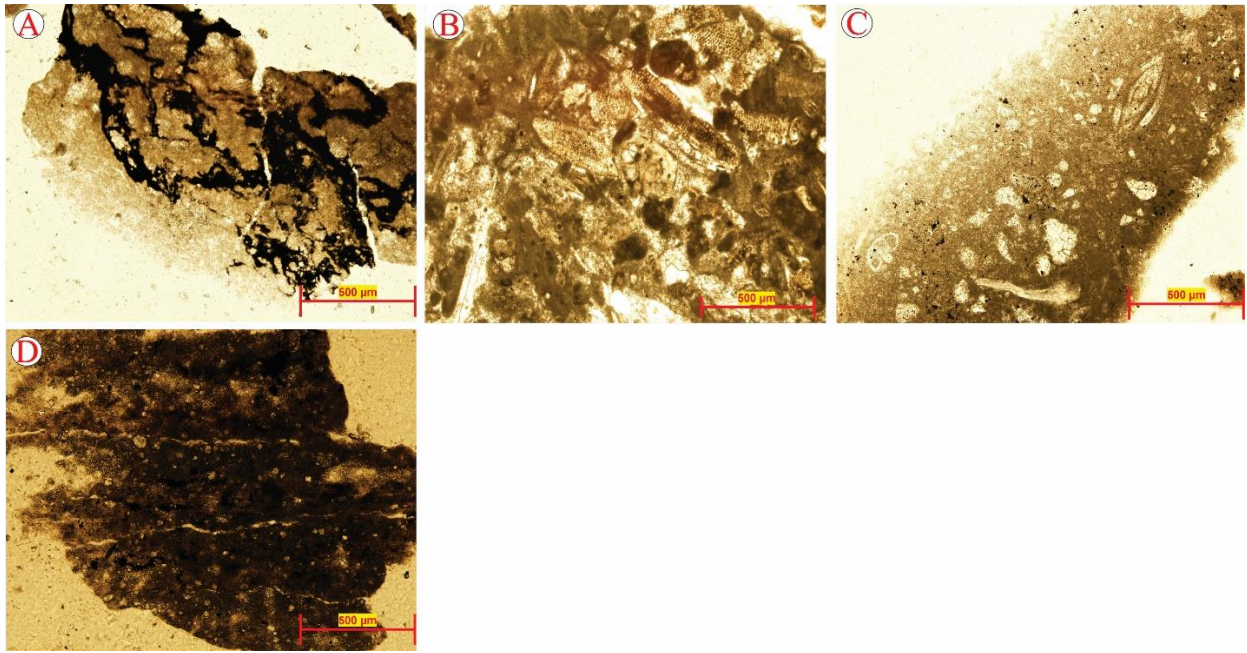

Fig. 4. (A) Mudstone (well D, PPL). (B) Pelloid echinoid bioclast wackestone to packstone (well A, PPL). (C) Benthic foraminifera bioclast wackestone (well C, PPL). (D) Shale petrofacies (well A, PPL).

## 4.2. Depositional models

Due to depositional sequence formation and its preservation in the stratigraphic columns, ancient sedimentary environments can be better analyzed and reconstructed (Selly, 1996). Facies sequences in carbonate formations are due to environmental changes. These changes may occur under the influence of natural internal processes such as intertidal zone regression, reefs regression, vertical aggregation of subtidal carbonate, movement of sand carbonate masses, and movements of shore-distant storms due to changes in the external controlling factors, including sea-level rise (Tucker and Wright, 1990). With regard to the previous studies, Alsherhan and Narin (1990) consider the Zagros basin and the Persian Gulf as an extensive platform that its northeast margin has existed since the Permian period. The Cretaceous period coincides with intense tectonic activities, including subduction of the Neo-Tethyan oceanic plate beneath the Iranian lithospheric plates during Early to Late Cretaceous time, emplacement ("obduction") of a number of Neo-Tethyan oceanic ophiolites over the Afro-Arabian passive continental margin in Late Cretaceous (Turonian to Campanian) time, and collision of the Afro-Arabian continental lithosphere with Iranian plates in Late Cretaceous and later times (Alavi, 2004). These types of Zagros deformations, along with the re-activation of basement faults, have a significant role in the evolution of the Zagros foreland basin during the Upper Cretaceous until Upper Miocene (Sherkati and Letouzey, 2004).

Identified facies in the Ilam Formation in wells located in Dezful Embayment and Abadan Plain indicate that carbonate sequences have been formed in a marine environment in which its depth and energy has constantly and periodically been changed. Therefore, in all four wells, shallow to deep facies have been identified. Based on the data and findings obtained from the investigation of lithofacies and their formation environments and considering the lack of extensive reef structures, lack of limestones containing slump structures or calciturbidite sediments in the Ilam Formation, a shallow carbonate platform of homocline ramp type can be suggested (Fig. 5, 6).

Overall, in all four studied wells, lagoon and shoal environments in the inner ramp settings have been identified, and no coastal facies and intertidal zone have been observed. The presence of foraminifera with a porcelaneous shell such as miliolid, and pelloid and the existence of limestone mud indicates a shallow and low-energy lagoonal sub-environment (Brandano, 2008). A gradual depth reduction, an increase in the environment's energy, a reduction in carbonate mud volume, and an increase of sparry calcite cement have occurred in the shoal facies belt located in the inner ramp, and thus, grainstone facies have been formed. Grainstone microfacies in all four wells have been identified. A rise in gamma and sonic logs across the stratigraphic column of the Ilam Formation can be due to shale petrofacies deposition. These shales had the characteristics of a shallow-depth environment, which suggests the sudden sea-level fall. Furthermore, facies observed from the four studies are related to the open marine located in the middle and outer ramps. They have characteristics of an abundance of planktonic foraminifera and oligostenginid fossils and an increase of mud carbonate. In figure (10), the stratigraphic column, along with the sedimentary environment of the Ilam Formation in the studied wells, have been depicted.

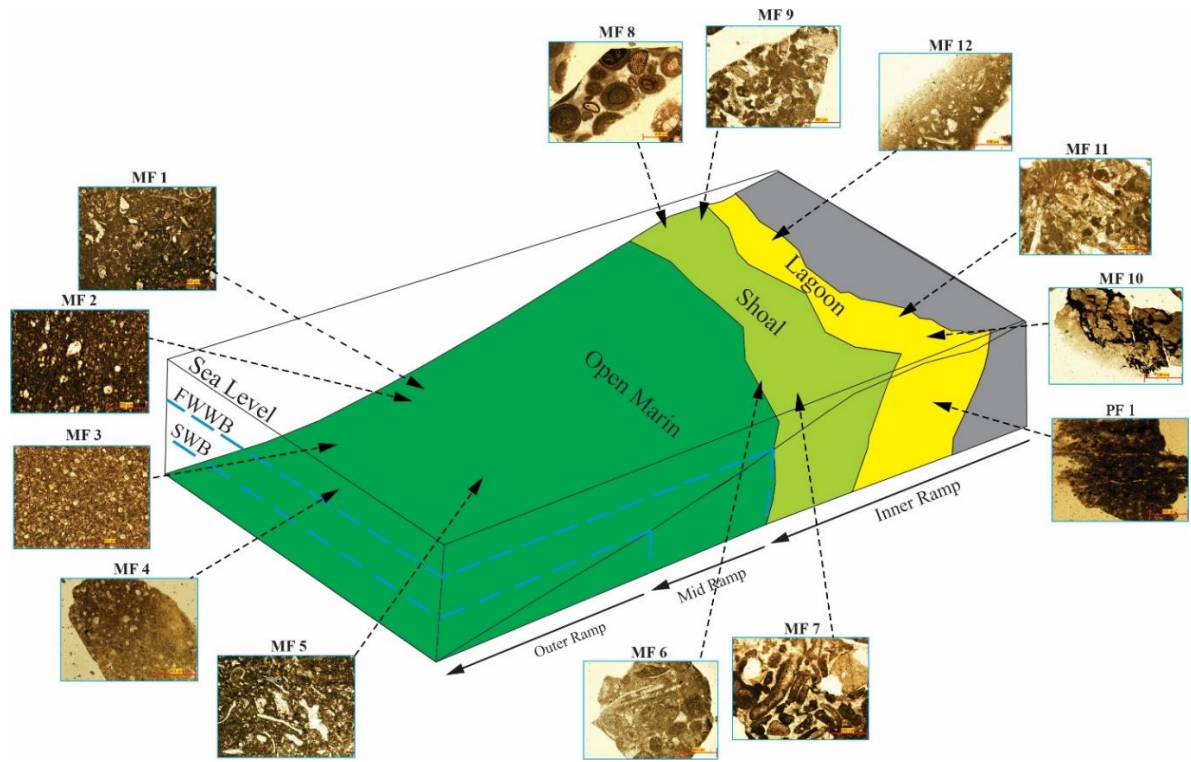

Fig. 5. Schematic image of the sedimentary environment of the Ilam Formation in the four studied wells located in the Dezful Embayment and Abadan Plain. The depositional model has been created according to Burchett and Wright (1992) and compared with the Flugel (2016) standards.

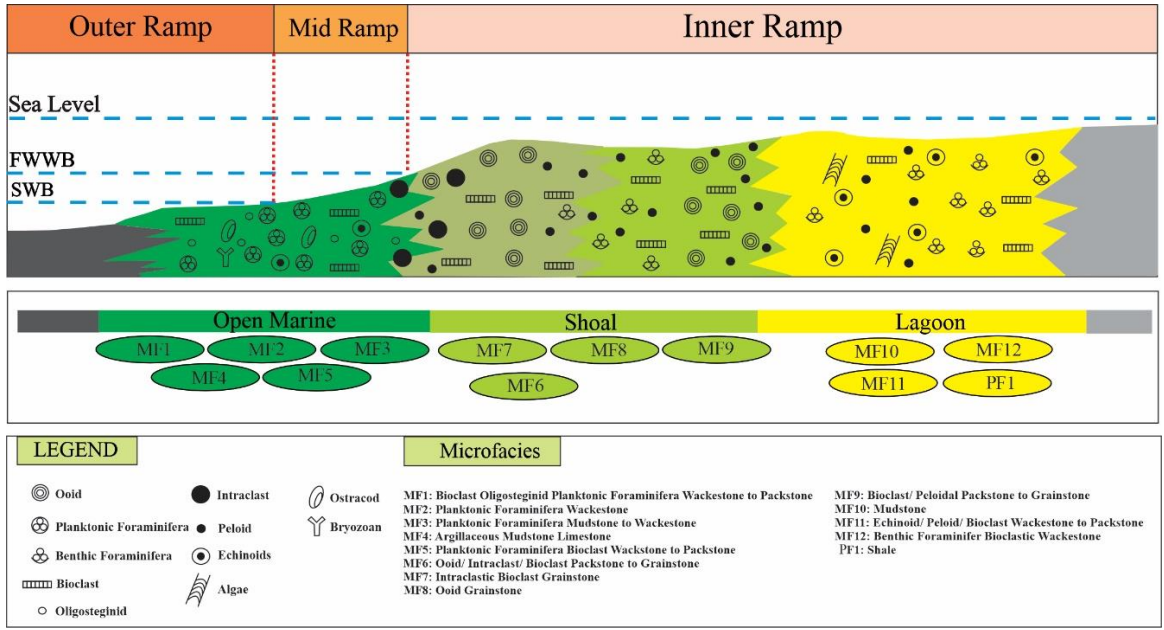

Fig. 6. Variation of sedimentary environment and facies belts in the Ilam Formation in the investigated wells.

## 4.3 Diagenetic processes

Diagenesis, referring to any physical or chemical changes in sediments or sedimentary rocks that occur after deposition (Scholle and Ulmer-Scholle, 2003)**.** Diagenetic processes are among the controlling factors of reservoir characteristics, and it is possible to predict reservoir quality through the study of diagenetic processes (Zhang et al., 2008). In fact, various diagenetic processes can influence the petrophysical characteristics, including total porosity, effective porosity, permeability, and pore size and its distribution, and produce zones with different petrophysical characteristics (Rahimpour- Bonab et al., 2013). Diagenetic processes in marine, meteoric, and burial environments have a significant role in the development and evolution of carbonate rock porosity (James and Choquettc, 1984). However, some diagenetic processes can reduce porosity and permeability (Heydari, 1997).

In this study, petrographic investigations show that cementation, dolomitization dissolution, compaction, and micritization are probably the main diagenetic processes affecting the pore space characteristics of the Ilam Formation in the studied wells.

### 4.3.1 Cementation

Several cement types are observed within the Ilam Formation. Drusy cement predominantly occurs as pore-filling and intergranular types (Fig. 7 A). Equant calcite cement is found inside the cavities and as intergranular cement in the form of mosaic (Fig. 7 B). Blocky cement occurs in veins, cavities, and as intergranular voids (Fig. 7 C). Turbid syntaxial cement is predominately observed in shallow facies and with a minor amount (about 10 percent) in open marine facies (Fig. 7 D). There are various types of fractures in varying sizes that are filled by disc-shaped, blocky, and drusy cement (Fig. 7 C). In addition, Fibrous isopachous cement has been formed around ooids and pelloids in grainstone facies (Fig. 7 E). Granular cement generally is observed in burial and meteoric environments. Recognition of the diagenetic environment of this cement is possible using a cathodoluminescence microscope and staining method (Fig. 7 G). Different generations of sparry calcite cement were reported in the Ilam limestone, ranging from marine through meteoric to some burial cements (Adabi and Asadi-Mehmandosti, 2008). These cements were possibly composed of aragonite, due to the identical morphology to that of recent warm-water shallow-marine aragonitic cements (Given and Wilkinson, 1985; Adabi and Rao, 1991).

### 4.3.2 Dolomitization

Dolomites are the minor diagenetic elements in the Ilam Formation and are as follows (Fig. 7 H, I). In the investigated sections, dolomite crystals have not been disconnected by stylolites which suggest that stylolites act as pore spaces for dolomite fluids and dolomitization has occurred after stylolitization. Dolomitization in some parts of the Ilam Formation in the studied section has occurred selectively. In some sections, only ooids became dolomitized, and in some other sections, only micritizied matrix became dolomitized, and it is observed in shallow-depth grainstone facies (Fig. 8 A, B).

### 4.3.3 Dissolution

Dissolution is one of the main diagenetic processes that often leads to the formation of secondary porosities such as moldic, vuggy, and channel porosities. Grains or allochems dissolve, and only their molds remain. Selective dissolution in aragonite type grains is found very extensively (Moore and Wade 2013). This porosity is often formed during burial and meteoric diagenetic types (Bathurst, 1975) (Fig. 8 D). Channel porosity is formed across fractures and cracks. These areas create a path for fluid movements. In this type of porosity, the length is much longer than their width and is often seen along with fractures (Tucker, 2001) (Fig. 8 E). Vuggy porosity is a fabric-independent porosity and is formed due to the dissolution of the rock parts with varying sizes. Vuggy porosity is irregular with a diameter greater than 1.16 mm, which is often unrelated to the rock fabric and is formed predominantly due to the dissolution of fabric-dependent pore spaces (Flugel, 2016) (Fig. 8 F).

### 4.3.4 Compaction

Both mechanical and chemical compactions are observed in the most studied intervals. This process in the studied subsurface samples occurred before cementation, and it led to the point, linear, and convex-concave contacts in the grains. Extension of physical compaction in the studied samples is often observed in the grain supported facies (Fig. 9 A, B, C). Across most of the stylolites, several insoluble residues, including organic materials, clay minerals, and iron minerals, are present. Furthermore, it is filled with iron or dolomite. Various types of peaks, high and low stylolites are found abundantly in the samples (Fig. 9 D). The abundance of chemical compaction in the form of dissolution seams (Fig. 9 F) and stylolites in the mud facies are predominantly higher and are observed differently in different parts of the Ilam Formation. The extent of this process in the upper part of Ilam, due to its grain supporting facies is smaller than the main and lower Ilam. The dominant portion of the main and lower Ilam is made up of mud facies, and thus, the compaction process is also higher in these facies. Stylolites are observed horizontally and parallel to the bedding, and peak amplitude ranges from less than a few millimeters to a few centimeters. The formation of dolomite crystals along the dissolution seams and stylolites is observed (Fig. 9 E). Some of the Dolomites in this formation are related to stylolite, which is a characteristic dolomitization type in the Middle Eastern Cretaceous carbonate sequences (Alsharhan and Nairn 1988).

### 4.3.5 Micritization

Most of the ooid grains are micritizied and are recognizable in grainstone facies. In some cases, extensive micritization led to the disappearance of skeletal structures and their transformation to pelloids (Fig. 9 G). Geopetal fabric of the lower part of the ostracode and the upper part of the cavities are filled by mud sediment and sparry calcite cement, respectively, during the diagenetic processes (Fig. 9 H). Bioturbation is present in some of the settings and has a darker color compared to the matrix. Borings are observed in the shallow and deeper part of the basin. (Fig. 9 I).

### 4.4 Pore types and characteristics

Investigations of carbonate rock porosity are significant in the understanding of diagenetic processes and reservoir analysis (Moore and Wade 2013). Choquette and Pray (1970) classified carbonate rocks based on their association or lack of association with rock fabric. The identified porosity types include intraparticle porosity (Fig. 8 C) and fracture porosity (Fig. 8 G). Fracture porosity is often a small percent of the total porosity in carbonate rocks. Open fractures increase the matrix permeability of reservoir rocks, which makes them crucial in terms of reservoir quality. (Nelson 1976; Rivas-Gomez et al. 2002; Lavrov 2017). Therefore, this type of porosity, due to connecting fractures together to increase permeability and hydrocarbon production, is important, and microporosity cannot be recognized by the naked eye (Fig. 8 H). Intercrystalline porosity, particularly in dolomite and sparry calcite crystal are basically secondary and show very low permeability (Fig. 8 I).

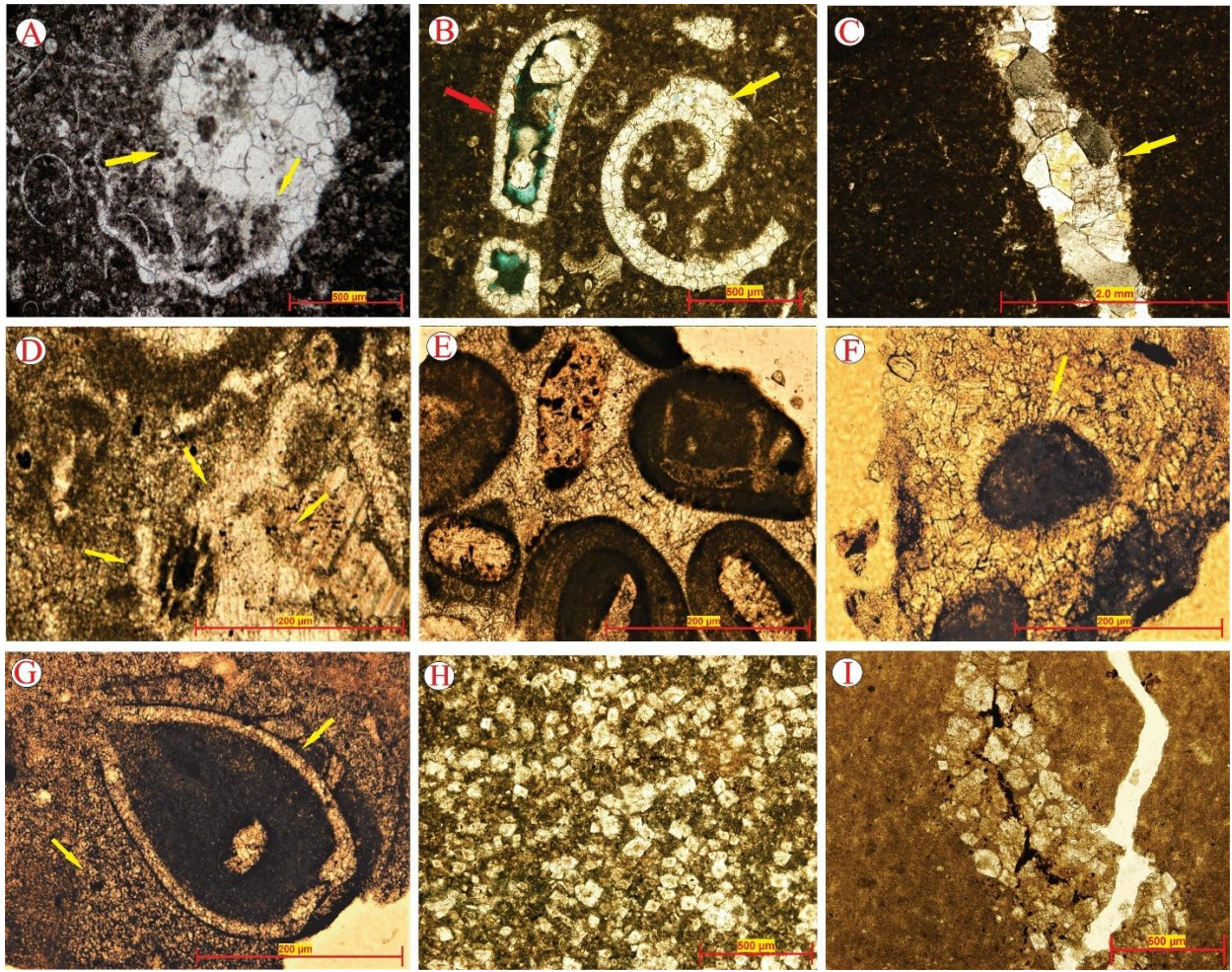

Fig. 7. (A) Image of a gastropod which is filled with drusy cement (well A, PPL). (B) Yellow arrow: Bioclast is filled by equant cement. Red arrow: A bioclast that has been dissolved and then filled by equant sparry (well A, PPL). (C) Vein filling cement: this vein is filled by blocky cement, which has a clear crystal boundary (well A, XPL). (D) Image of possibly turbid syntaxial marine cement around crinoid (well D, PPL). (E) Fibrous isopachous cement around ooids in grainstone facies (well D, PPL). (F) Bladed cement around a pelloid (well B, PPL). (G) Bivalves shell is filled with granular cement (well D, PPL). (H) Dolomicrosparite observed in the sediments of the Ilam Formation. This dolomite replaced mud matrix dolomites have clear rim and cloudy center (well A, PPL). (I) It is traces of hydrocarbon within dolomite rhombs (well A, PPL).

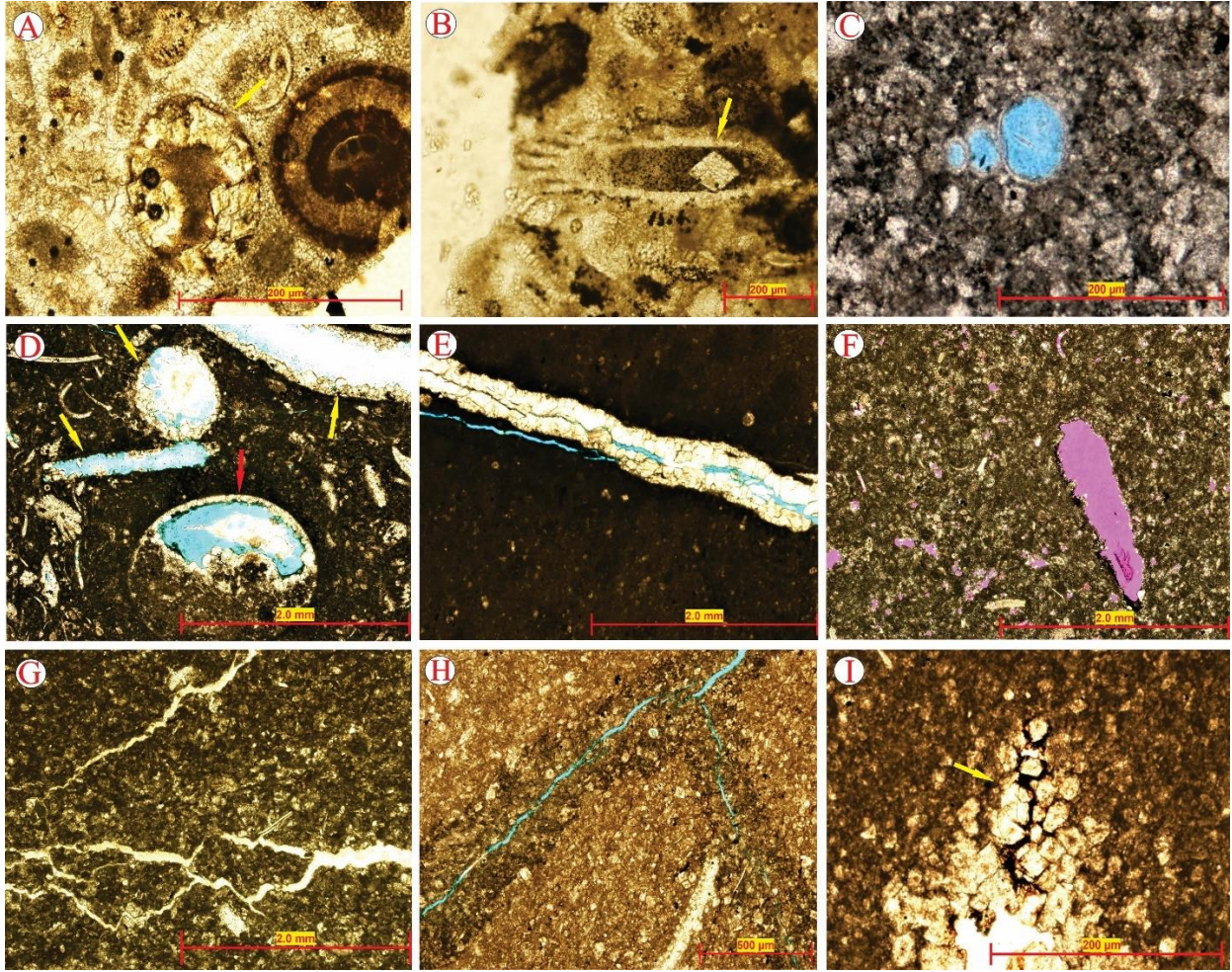

Fig. 8. (A) Image of selective dolomitization in ooids of the Ilam Formation. The selective dolomitization is not intense and ooids are easily recognizable (well A, PPL). (B) Rhombohedron dolomite within an echinoid (well A, PPL). (C) Intraparticle porosity within planktonic foraminifera (well A, PPL). (D) Yellow arrow: bioclast debris which has been dissolved under the influence of meteoric processes and caused moldic porosity. Red arrow: A geopetal fabric in a shell debris, with mud in the lower part and sparry calcite cement in the upper part (well A, PPL). (E) Channel porosity that has been reduced by drusy cement (well A, PPL). (F) Vuggy porosity. (well A, PPL). (G) Image of porosity caused by multi fractures in a mudstone facies (well A, PPL). (H) Fracture porosity, along with a number of microporosities in the matrix have been formed. (well A, PPL). (I) Interacrystaline porosity and hydrocarbon materials within the dolomite's crystals (well A, PPL).

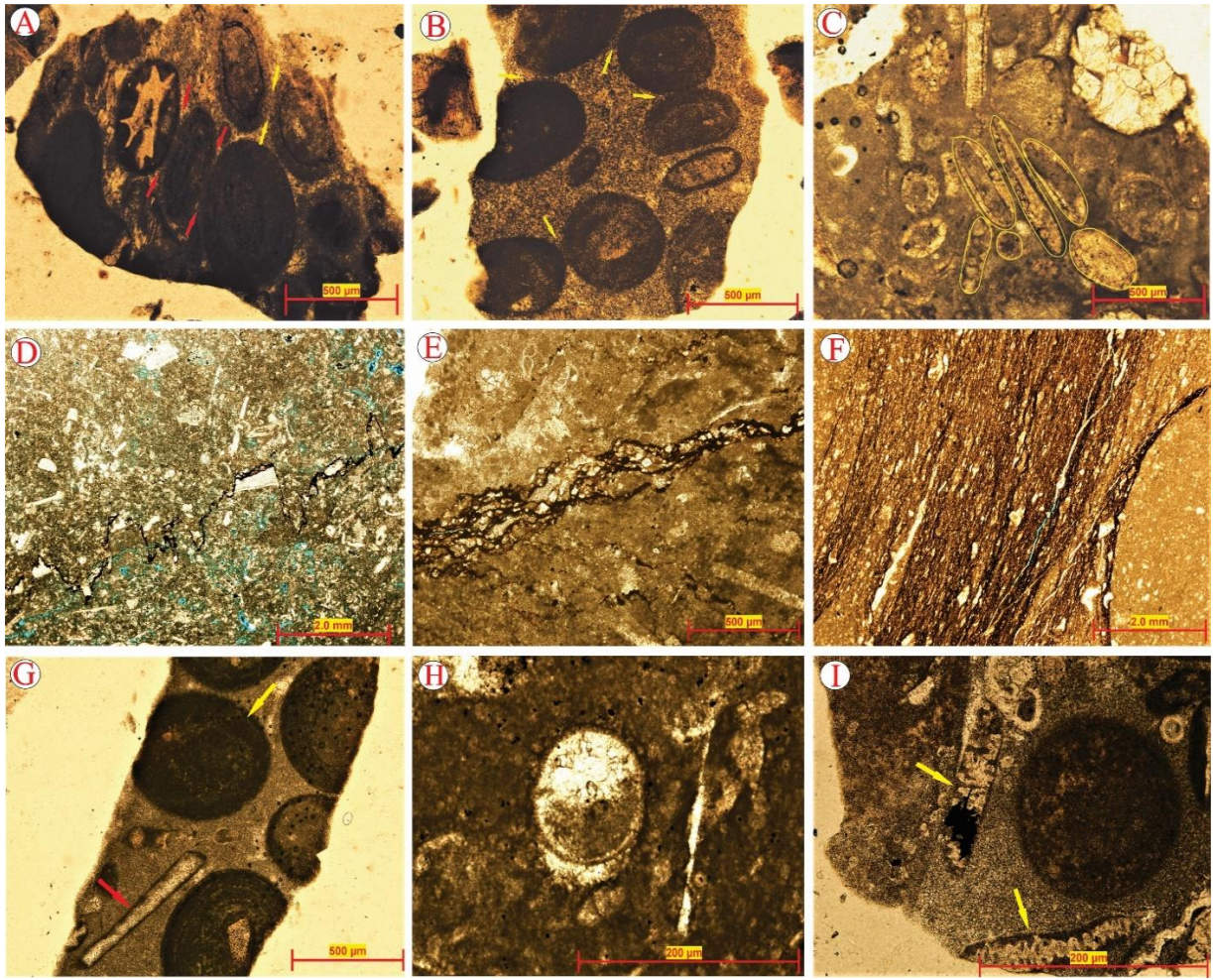

Fig. 9. (A) Linear contacts between micritized ooids due to physical compaction. (well D, PPL).
(B) Point contact between ooids due to physical compaction. (well D, PPL). (C) Orientation of
ooids due to mechanical compression (well A, PPL). (D) High amplitude stylolites. (well A, PPL).
(E) Dissolution seams with a few dolomite crystals (well A, PPL). (F) These dissolution structures
are formed due to the chemical compaction around a primary carbonate nodule. The chemical
compaction led to the formation of a large number of dissolution cracks or microstylolites. (well
A). (G) Micritization of ooids in oomicrite facies (yellow arrow) forming bahamite. Cortoid or
micritic envelope around fossil debris (red arrow). (well D, PPL). (H) Geopetal fabric mud is in
the lower part, and sparry calcite is in the upper part (well A, XPL). (I) Boring by endolithic algae
or cyanobacteria. (well D, PPL).

## 4.5 Sequence stratigraphy

In sequence stratigraphy, sedimentary bodies are defined and interpreted based on their stratal
stacking patterns and their stratigraphic relationships (Catuneanu, 2017) in combination with the
occurrence of unconformity surfaces of varying orders and their correlative conformities
(Catuneanu, 2006). Eustatic sea-level changes and/or tectonic processes at different temporal and
spatial scales may impact the variations in sequence development (Brunet et al., 2009).

Results obtained from petrographic investigations of the Ilam Formation's facies, along with analysis of the petrographic logs in the studied wells, indicate that the Ilam Formation, which is Cenomanian–Santonian in age (James and Wynd, 1965) consists of a third-order sedimentary sequence. This sedimentary sequence in all four studied wells has a varying thickness ranging from 106 to 141 meters. In the following section, this identified sedimentary sequence will be explained. To determine depositional sequences and sequence boundaries, the sedimentary models of Hunt and Tucker (1992) were used.

### 4.5.1 Sequence stratigraphic investigations in Wells

The petrographic investigations of the studied sedimentary sequence show no evidence of a sequence boundary at the base of the Ilam Formation. However, there was a recognizable unconformity at the base of the Laffan member, which is probably the sequence boundary. A sequence boundary is recorded in all of the studied wells, where the deep marine deposits of the Gurpi Formation is observed. Considering the fact that the age of this sequence is Coniacian-Santonian, therefore, it is regarded as a third-order sedimentary sequence. Based on the petrographic investigations, facies variations, and the extent of gamma log changes across the stratigraphic column of this formation, two types of system tracts, i.e., transgressive systems tract (TST) and highstand systems tract (HST), were recognized in this formation (Fig. 10). TST, based on the petrographic investigations, has an upward deepening trend and is made up open marine facies which contain planktonic foraminifera (heterohelix, hedbergella, and globigerinelloides) with a mud supporting matrix. The variation curve of sea level in this part of the studied sequence indicates a low-energy environment. The Gamma log indicates a small amount of clay in this part of the formation. Due to the extensive seawater transgression, the relative seawater depth reached to it's maximum, and the maximum flooding surface (MFS) with the sudden decline of gamma log was recognized. From this time onward sea-level rate slowed down, and sedimentation on the carbonate platform led to the formation of HST, and shallowing upward trend occurred. Following MFS, there was a gradual change from deep waters to inner ramp facies, the high-energy shoal and low-energy lagoonal environments. The upper boundary of this sequence belongs to the lagoonal environment and is considered a sequence boundary.

Towards North-East of the studied area (Well No. D), due to the deep marine deposits, the thickness of the wells and TST increases. There is no other significant changes observed in this correlation chart (Fig. 10).

Fig. 10. Correlation chart of the sequences stratigraphy of the Ilam Formation in wells No. A to D.

Table. 1. Existing facies in TST and HST in wells No. A to D.

| | TST | HST |
|---|---|---|
| well No. A | MF1, MF2, MF3, MF12 | MF5, MF12, PF1, MF9, MF7, MF8, MF6, MF11 |
| well No. B | MF4, MF9, MF3 | PF1, MF12, MF8 |
| well No. C | MF3 | MF6, MF8, MF10, MF4, PF1, MF12 |
| well No. D | MF9, MF3 | MF6, MF8, MF10, MF9, PF1, MF12 |

## 4.6 Reservoir rock classification

Three different reservoir rock classification methods were used in this study to construct and verify the final field-scale reservoir zonation (see also supplementary material for more details of the methods).Using core data, logs, and thin sections, the reservoir rock analysis of the Ilam Formation in the subsurface of one of the studied wells was investigated based on the following methods.

### 4.6.1 Carman and Kozeny Equation

In the oil industry, core analysis and well testing is a standard method for permeability determination. On the other hand, there are several theoretic, experimental, and mathematical relationships for estimating permeability. Among them Carman Kozney relation is the most important and industrially well-known equation to estimate permeability and can be expressed as equational:

$$k = \frac{\phi_e^3}{(1 - \phi_e)^2} \left[ \frac{1}{2\tau^2 s_{gv}^2} \right] \tag{1}$$

Where K: permeability, $\phi_e$: effective porosity, $\tau$: tortuosity, $S_{gv:}$ special grain volume

In the current study, two methods were employed, which are both based on simple mathematical manipulation of Carman – Kozeny: Amefule's method and Gunter's method. A more detailed description of the mentioned methods can be found in the following:

### 4.6.2 Amaefule et al. method

A flow unit is a volume of reservoir rock that is vertical and lateral, constant, and predictable, and the geological and petrophysical characteristics that affect the flow of fluid are constant within it and are distinguishably different from other rock volumes (Tiab and Donaldson, 2004).

Flow units can be determined using various methods. In Amaefule et al. method (1993), each flow unit is marked with a flow zone indicator (FZI), and the flow zone indicator is a function of the reservoir quality indicator (RQI). The calculation of the flow zone indicator and reservoir quality indicator is based on porosity and permeability cores. Hydraulic flow unit is based on the relationship between permeability and porosity and is basically proposed by Kozeny (1927) and Carman (1937). Kozeny-Carman is a theoretical method for the dependence of permeability on porosity structure (Amaefule et al., 1993). Equation 2 is as follows:

$$\phi_z = \frac{\phi_e}{1 - \phi_e} \tag{2}$$

In Equation 2, $\phi_z$ is normalized porosity that represents hole volume to grain volume ratio, $\phi_e$: effective porosity.

$$RQI = \sqrt{K/\phi_e} \tag{3}$$

Equation 3 RQI: reservoir quality indicator, K permeability (millidarcy unit), $\phi_e$: effective porosity.

In Equation 3, RQI represents the reservoir quality indicator. This indicator is an approximation of reservoir rock and hydraulic radius average. It is a key for hydraulic units that are associated with porosity, permeability, and capillary pressure to each other (Tiab and Donaldson, 2015).

$$FZI = \frac{RQI}{\phi_z} \tag{4}$$

Equation 4: $\phi_z$ normalized porosity, RQI Reservoir quality indicator, FZI flow zone indicator.

Rocks containing a limited amount of FZI belong to a single flow unit, which means they have the same flow characteristics (Prasad, 2003). Methods for determining the number of flow units include normal probability diagram, clustering methods, and histogram analysis. FZI data clustering is a proper method for determining the optimal number of flow units.

### 4.6.3 Gunter method based on the Winland equation and Lorenz method

Through integrating porosity, permeability, and water saturation with capillary pressure mercury injection data, Winland (1972) obtained an experimental equation among porosity, air permeability, and pore spaces related to 35 percent mercury saturation which became a basis for numerous future studies. The Winland method establishes an association among porosity, air permeability, and pore spaces related to 35 percent mercury saturation. This method can be used for other percentages (30, 40. 50). However, 35 percent has the most precision (Al-Qenae and Al-Thaqafi, 2015). R35 indicator is the bottleneck radius of the calculated pore spaces in 35 percent mercury saturation in permeability, porosity, and injection mercury capillary pressure test that can be calculated via equation 5 (Winland equation).

$$log(R35) = 0 \cdot 732 + 0 \cdot 588 \, log(K) - 0 \cdot 864 log \, (\varphi) \tag{5}$$

In this method, R35 value for the same rock types is equal. Therefore, rock typing can be done with the consideration of the created areas and inserting the porosity-permeability data on the Winland diagram. These types indicate the same and predictable flow characteristics. In this procedure, permeability is calculated for R35 value, and different porosities and zonation will be created (Porras and Campos, 2001).

Different areas of Winland analysis are determined based on the size of bottleneck radius, which in this study are as follows:

- Units with R35 smaller than 0/1 micron
- Unites with R35 between 0/1 and 0/5 micron
- Unites with R35 between 0/5 and 2 microns
- Unites with R35 between 2 and 10 microns

Winland method for determining the areas is as follows:

- Calculation of R35 for all the porosity and permeability data
- Ascending ordering of the calculated R35 data
- Drawing permeability semi-log plot based on porosity with the consideration of identical porosity lines (Al-Qenae and Thaqafi, 2015).

For confirmation of the results obtained from the Winland diagram, the SMPL diagram or the stratigraphic modified Lorenz plot (SML) was employed.

### 4.6.4 Lorenz method (SMLP)

The stratigraphic modified Lorenz plot (SML) is one of the best methods for obtaining the minimum number of flow units in a reservoir (Gunter et al., 1997). The Lorenz method (SMLP) is based on total flow capacity with maintaining stratigraphic order. The inflection points on the plot represent variations in the flow characteristics of a porous environment. For determining flow units in the Lorenz method, in the first step, continuous permeability and porosity and permeability to porosity ratio ($K/\phi$) in a stratigraphic arrangement are ordered. Following that, the product of permeability in related depth (k.h) and porosity in related depths ($\phi$.h) is calculated. The cumulative sum of the data obtained from the multiplication of permeability and depth, along with multiplications of porosity and depth, is calculated. As the next step, the data is normalized to 100 percent. The obtained data from the multiplication of permeability and depth is called flow capacity. The obtained data from the multiplication of porosity and depth is called storage capacity. Subsequently, storage capacity values and flow capacity are plotted (Gomes et at., 2008).

In a stratigraphic order, recognizable flow units include reservoir units, flow barriers or bars, speed zones, and bafels or zones that cause turbulence in the movement of fluids. The characteristic of these key units is that the reservoir units have high flow and storage capacities, and the speed unit has a high flow capacity and very low storage capacity. Bafel Zones have a high storage capacity but low flow capacity, and ultimately barrier zones have very low or almost zero flow and storage capacities (Gunter et al., 1997).

### 4.6.5 Procedure

In this study, a fully automated approach designed and proposed by (Ghasemi, Kakemem, et al., 2022) was employed for the lower part of the Ilam Formation in one of the studied wells. Amaefule et al. (1993) method was utilized to determine the flow units with flow zone indicator (FZI) and reservoir quality indicator (RQI). These calculations have been conducted using equation 2 and equation 3 for the calculation of RQI and equation 4 for the calculation of FZI. Following this, FZI data were clustered so as to determine the optimal number of flow units using the MATLAB program (Fig. 11) and (Fig. 12). The K-means algorithm was employed for the purpose of clustering. However, due to the fact that the number of clusters has been selected manually in K-means and to overcome this issue in this study, the Elbow method was used for determining the optimal number of clusters (Fig. 13). The result of the Elbow method suggests 6 clusters to be generated by the K-means algorithm in order to obtain the finest result. As a result, the FZI value is divided into 6 clusters. Based on the Amaefule et al. (1993) method, six types of flow units were obtained. If we place RQI values on $\phi z$ on a log-log sheet using Amaefule et al. (1993) method and the dots are separated, those dots that have been plotted on lines with a slope of 1 and with an equal intercept represent one flow unit. As is seen in the figure, the number of these lines is the same as clustered FZI (Fig. 14).

The first step in the Gunter method is to determine the R35 data and then cluster them using K-means and Elbow methods. In this method, each flow unit that has an equal R35 is separated. Using the Winland method, the flow units were classified based on the R35 values, and thus, there are six flow units in this method (Fig. 15) and (Fig. 16). For separation of the R35 values using the Winland diagram, the cross plot of K was drawn based on $\phi$ (Fig. 17). The next step is drawing the stratigraphic modified Lorenz plot (SML), which is one of the best methods for obtaining the minimum number of flow units in a reservoir. For determining flow units in the Lorenz method, firstly, continuous permeability and porosity and permeability to porosity ratio (K/$\phi$) in a stratigraphic arrangement are ordered. Following that, the product of permeability and related depth (k.h) and porosity and related depths ($\phi$.h) is calculated. The cumulative sum of the data obtained from the multiplication of permeability and depth and porosity and depth is calculated, and then the data is normalized to 100 percent (Fig. 20). Each slope represents a flow unit. Therefore, in the first stage, the variations of the slopes were determined. The PELT algorithm determines the sudden variation of slopes; however, this algorithm requires that the dots having slope variation be determined manually. Hence, to overcome this problem, the Elbow method was used, resulting in 8 as the optimal number of segments (Fig. 18) and (Fig. 19). However, the optimal number of segments is not necessarily the same as the number of flow zones that was distinguished by R35 and FZI since it doesn't determine this number of segments having variations and it might observe the same slope in two or three places on the plot. Given this issue, every two equal slopes represent one flow unit, and with regard to the drawn plot, six flow zones were determined (Fig. 21).

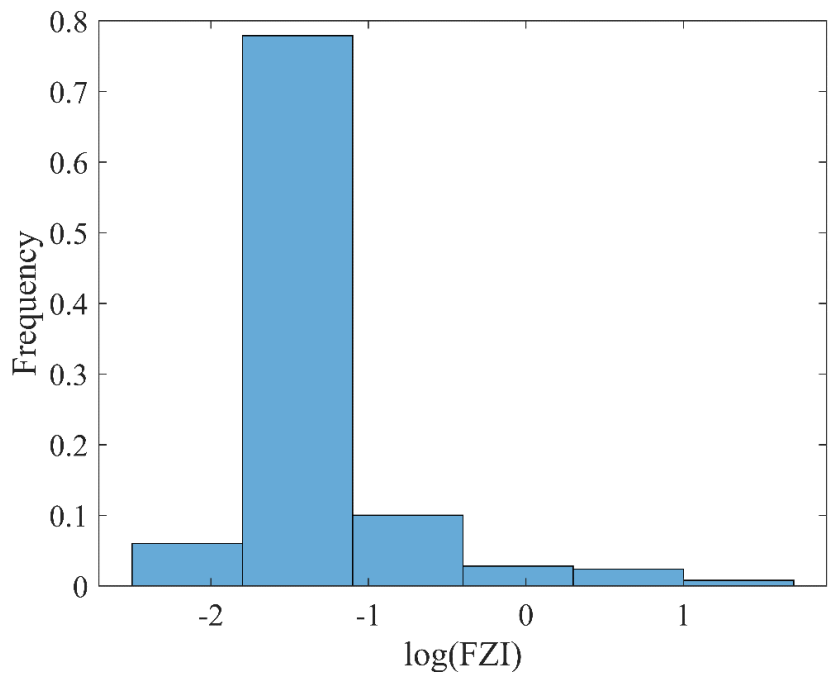

Fig. 11. Histogram of logarithm of FZI. The number of bars is equal to the optimum FZI clusters number.

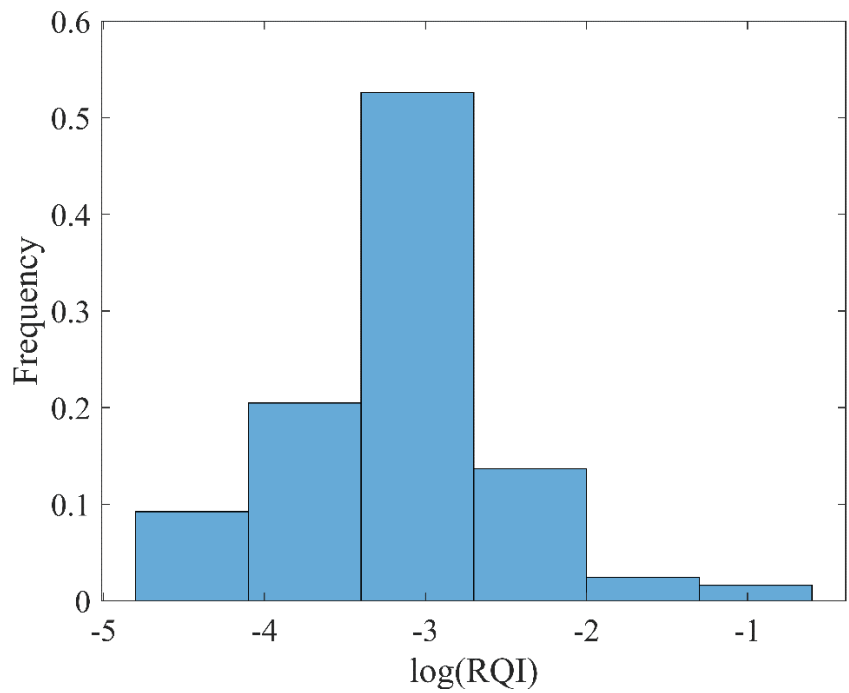

Fig. 12. The reservoir quality indicator (RQI) frequency distribution histogram using the Amaefule method.

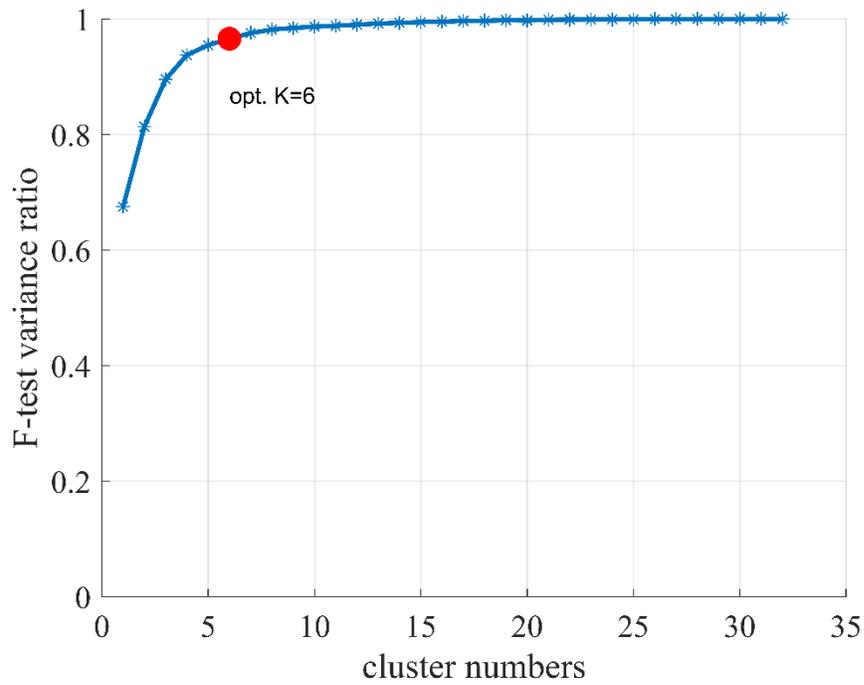

Fig. 13. Elbow diagram computed for logarithm of flow zone indicator (FZI) values.

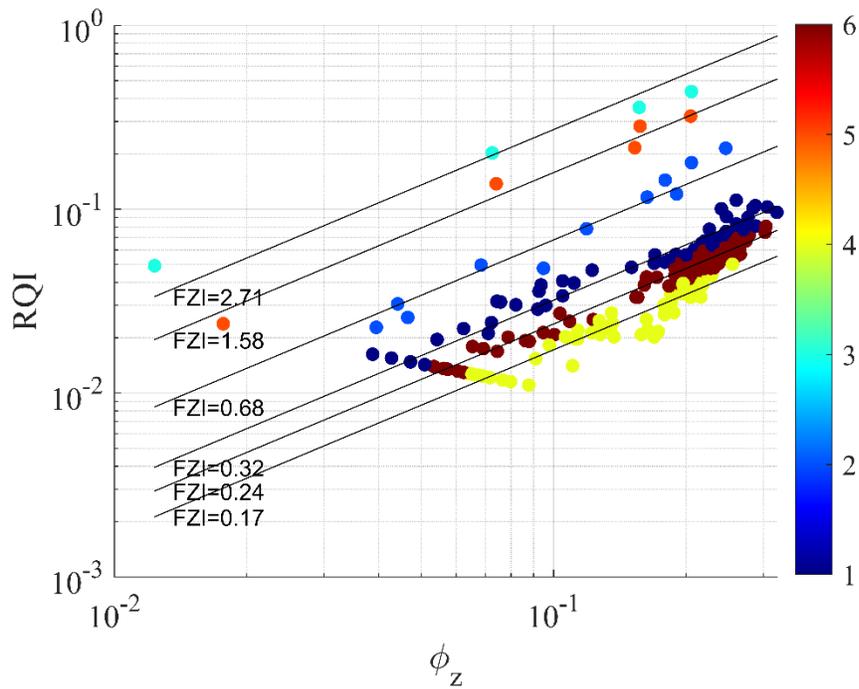

Fig. 14. Amaefule's FZI diagram. The number of clusters was determined by the Elbow algorithm.

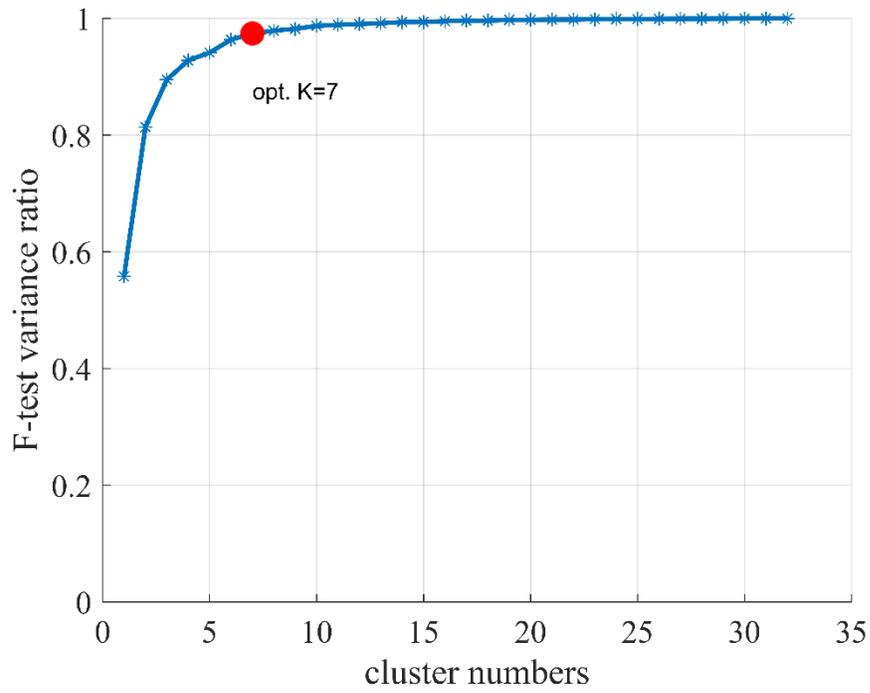

Fig. 15. The diagram of the optimal number of R35 clusters using the Elbow method.

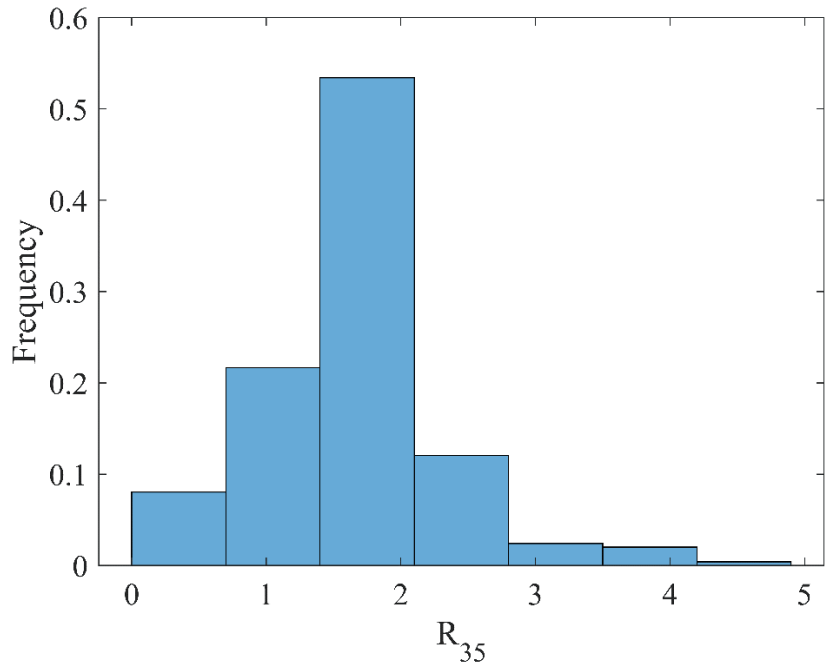

Fig. 16. The R35 frequency distribution diagram using the Winland method.

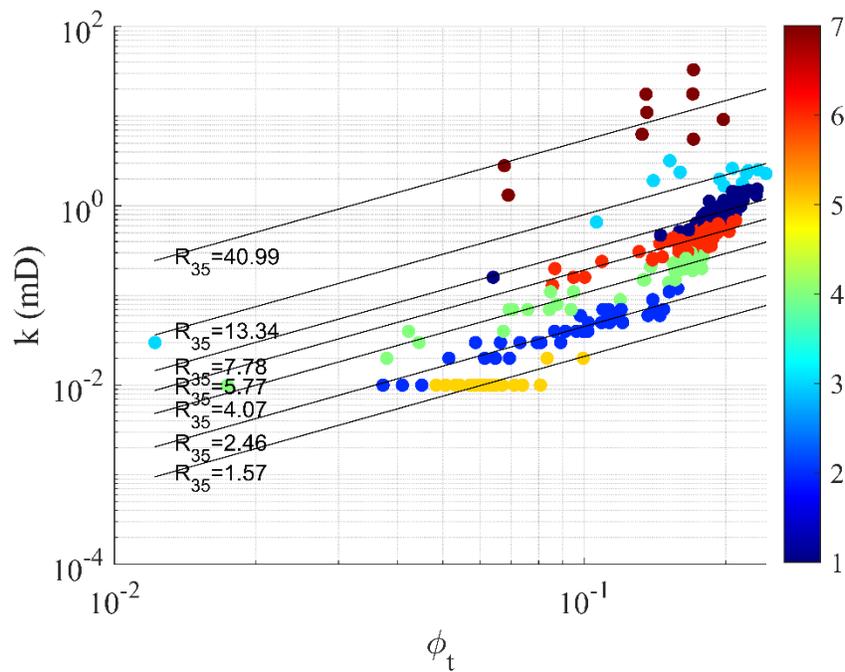

Fig. 17. The separation of R35 values using the Winland diagram.

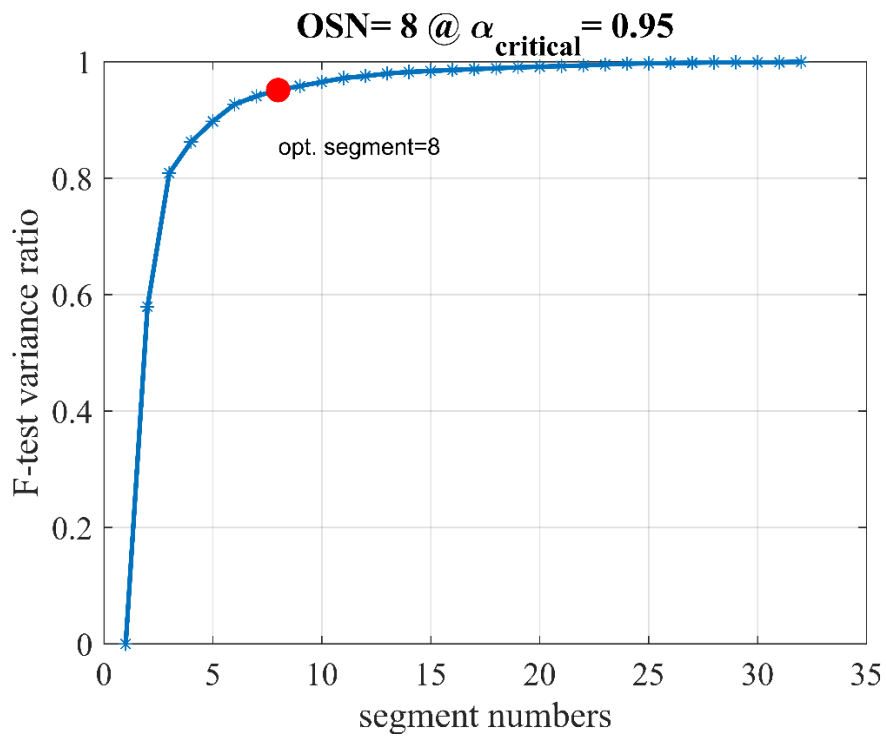

Fig. 18. The optimal number of segments using the Elbow method.

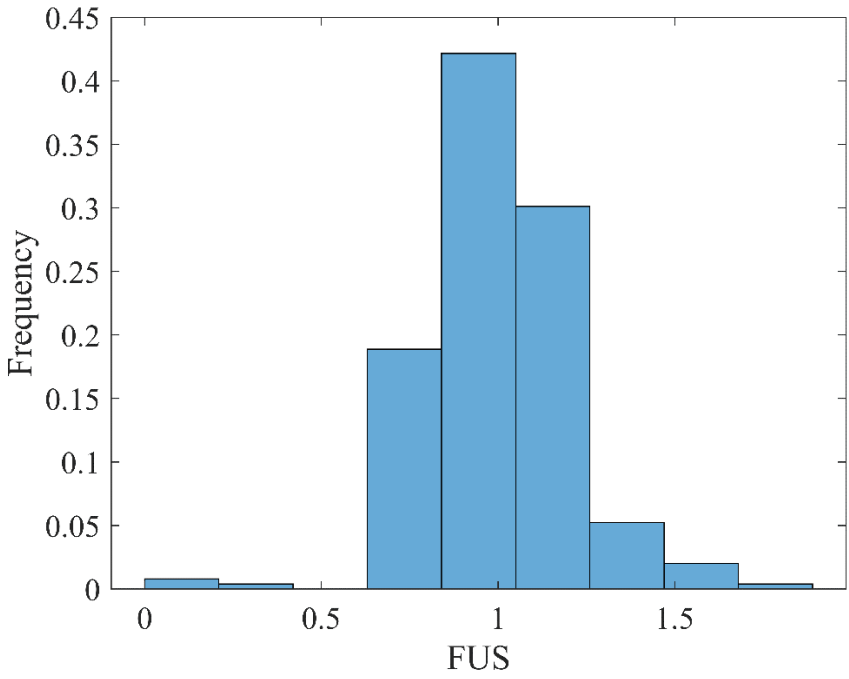

Fig. 19. Histogram of FUS frequency distribution.

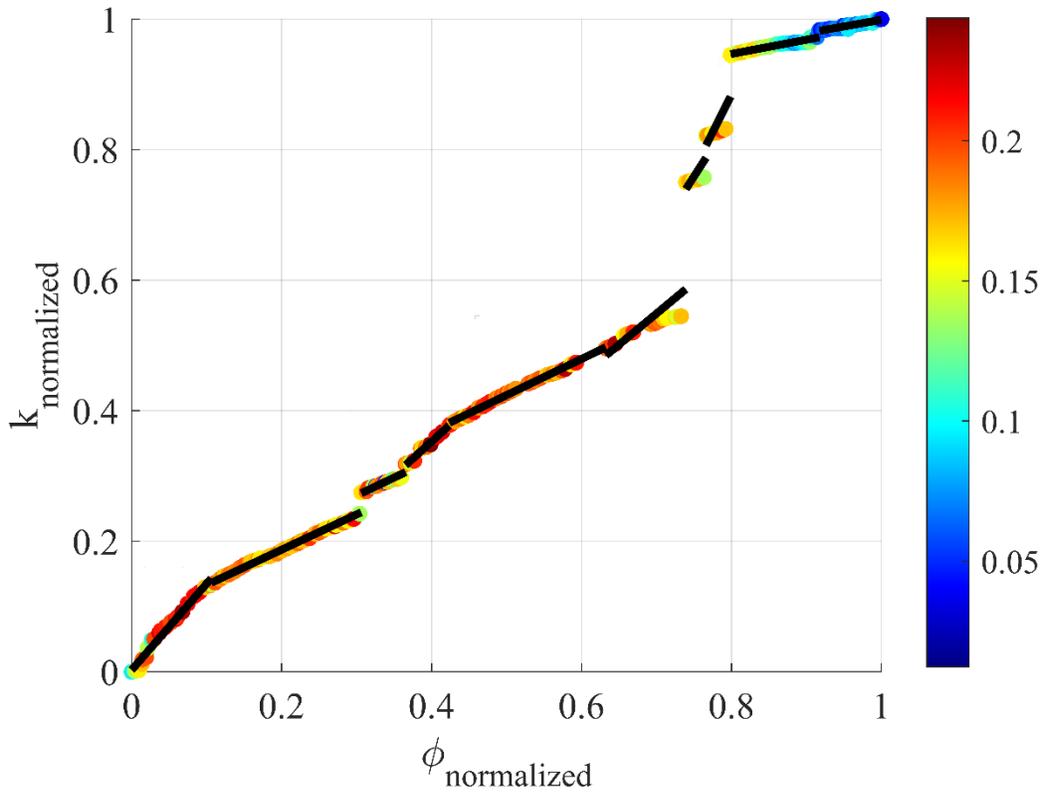

Fig. 20. The stratigraphic modified Lorenz plot (SMLP).

## 4.6.6 Interpretation of flow units

**Flow unit 1:** This flow unit begins at the depth of 2850.13 meters and ends at the depth of 2855.74 meters. The lithology of this part is limestone which contains MF3 and MF5 microfacies, respectively. FZI in this flow unit is equal to 0.384. The permeability range is equal to 1.764 millidarcy and the porosity range is equal to 0.185 (Table 2). The sedimentary environment of this part belongs to open marine. Intense dissolution, cementation, pyritization, chemical compaction, micritization, and bioturbation are among the diagenetic processes in this part of the sequence. This flow unit due to intensive dissolution has fracture, channel, vuggy, interaparticle, moldic porosities, and porosity in the matrix. This part was measured based on HFU and belongs to Bafel unit in terms of reservoir quality (Fig. 21).

**Flow unit 2:** this flow unit begins at the depth of 2855.55 meters to 2869.48 meters and it can be found even from 2872.55 meters to 2880.84 meters. According to the petrographic investigations, this part has a lithology of limestone which contains MF12, MF3, and MF1 microfacies. FZI in this flow unit is equal to 0.277. The permeability range is equal to 0.900 millidarcy and the porosity range is equal to 0.185 (Table 2). The sedimentary environment in this part of sequence belongs to open marine setting. Dolomitization especially around stylolites, formation of vein calcite cement in fracture, cementation in foraminifera chambers, micritization, and chemical compaction are among the diagenetic processes in this part. In addition, porosities of fracture, channel, and interaparticle types as well as vuggy porosity can be observed to some extent. This part was measured based on HFU and belongs to Bafel unit in terms of reservoir quality (Fig. 21).

**Flow unit 3:** This flow unit begins at the depth of 2869.83 meters to 2872.17 meters and it can be found even from 2882.57 meters to 2887.55 meters. According to the petrographic investigations, this part has a lithology of limestone which contains MF3 microfacies. FZI in this flow unit is equal to 0.277. The permeability range is equal to 0.636 millidarcy and the porosity range is equal to 0.171 (Table 2). The sedimentary environment of this part of sequence belongs to open marine setting. Dolomitization, micritization, pyritization, bioturbation, cementation, interaparticle, and chemical compaction are among the diagenetic processes in this part. This part was measured based on HFU and belongs to Bafel unit in terms of reservoir quality (Fig. 21).

**Flow unit 4:** This flow unit begins at the depth of 2887.89 meters and ends at the depth of 2889.21 meters. According to the petrographic investigations, this part of sequence has a lithology of limestone which contains MF3 microfacies. FZI in this flow unit is equal to 0.315. The permeability range is equal to 2.128 millidarcy and the porosity range is equal to 0.180 (Table 2). The sedimentary environment of this part belongs to open marine sub-environment. Dolomitization, micritization, pyritization, bioturbation, and chemical compaction are among the diagenetic processes in this part. This part was measured based on HFU and belongs to Bafel unit in terms of reservoir quality (Fig. 21).

**Flow unit 5:** This flow unit begins at the depth of 2889.41 meters and ends at the depth of 2890.79 meters. According to the petrographic investigations, this part of the sequence has a lithology of limestone which contains MF3 microfacies. FZI in this flow unit is equal to 0.491. The permeability range is equal to 2.746 millidarcy and the porosity range is equal to 0.160 (Table 2). The sedimentary environment of this part of sequence belongs to open marine sub-environment. Dolomitization, micritization, pyritization, bioturbation, cementation, and chemical compaction are among the diagenetic processes in this part. In addition, porosities of fracture as well as microporosity in the matrix can be observed. Although a number of fractures have been filled with cement, the presence of microporosities in the matrix and fractures without cement resulted in a permeability increase in this unit. This part was measured based on HFU and belongs to the Speed zone unit. Therefore, flow unit 5 contains the best reservoir quality among the flow units (Fig. 21).

**Flow unit 6:** This flow unit begins at the depth of 2991.14 meters and ends at the depth of 2911.23 meters. This part contains limestone and MF3 microfacies. FZI in this flow unit is equal to 0.374. The permeability range is equal to 0.212 millidarcy and the porosity range is equal to 0.092 (Table 2). The sedimentary environment of this part of sequence belongs to the open marine setting. Micritization, bioturbation, and chemical compaction, and low porosity are among the diagenetic processes in this part. This part was measured based on HFU and belongs to the Barrier unit. Therefore, flow unit 6 contains the worst reservoir quality among the flow units (Fig. 21).

Table 2. Measured mean values (permeability (K), porosity (φ), reservoir quality indicator (RQI), flow unit indicator (FZI), and HFU) of each flow unit.

| | $K_h$ | $\varphi_h$ | $RQI_h$ | $R35_h$ | $FZI_h$ | $HFU_h$ |
|---|---|---|---|---|---|---|
| Flow unit 1 | 1.746 | 0.185 | 0.086 | 9.831 | 0.384 | 1.323 |
| Flow unit 2 | 0.900 | 0.185 | 0.063 | 6.930 | 0.277 | 0.547 |
| Flow unit 3 | 0.636 | 0.171 | 0.054 | 5.829 | 0.259 | 1 |
| Flow unit 4 | 2.128 | 0.180 | 0.069 | 6.139 | 0.315 | 1.785 |
| Flow unit 5 | 2.746 | 0.160 | 0.084 | 6.093 | 0.491 | 2.291 |
| Flow unit 6 | 0.212 | 0.092 | 0.031 | 3.16 | 0.374 | 0.210 |

Based on the investigations and obtained data, flow unit 5 has the best reservoir quality, and flow unit 6 is the worst (Fig. 21). Although unit 5 has the same microfacies and almost equal amount of porosity as unit 3, possesses a better reservoir quality, which is probably due to its diagenetic processes (microporosities in the matrix and porosities caused by fracture and effective porosity) that increased permeability in this depth.

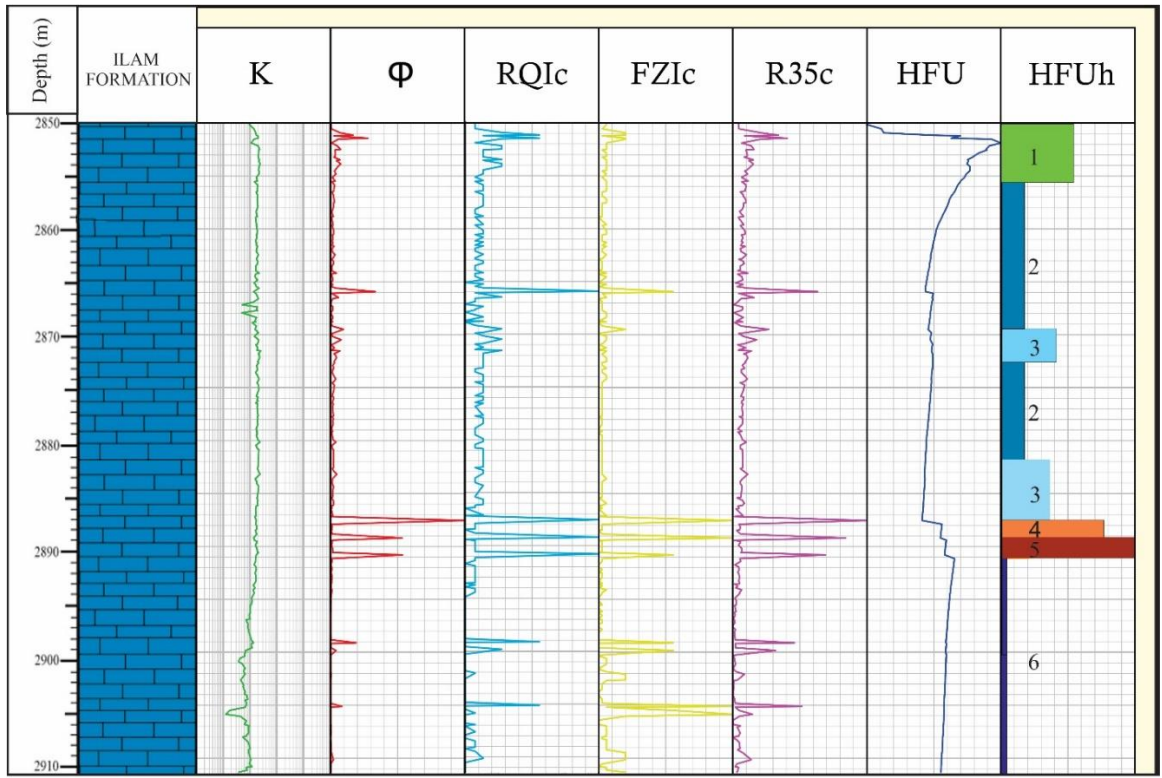

Fig. 21. Flow units along with stratigraphic column of the Ilam Formation at the studied depth.

## 5. Conclusion

Based on the study of the depositions of sequences of the Ilam Formation (Cenomanian–Santonian in age) in four subsurface sections located the Dezful Embayment and Abadan Plain in southwest Iran, the following results were obtained. Lithology of the Ilam Formation in the studied wells is mainly limestone with interbedded shale and argillaceous limestone. The upper contact is with the Gurpi Formation and the lower contact is with Laffan member.

(1) Based on the petrographic investigations on the Ilam Formation, 12 microfacies and one shale petrofacies were recognized. They have been deposited in the lagoon, shoal, and open marine facies belts.

(2) The recognized microfacies suggest that the sedimentary environment of the Ilam Formation can be considered as a shallow water carbonate platform of homoclinal ramp type settings and sediments have been deposited in several sub-environments.

(3) The results obtained from the petrographic investigations on the facies of the Ilam Formation along with the petrographic logs indicate that it is composed of a third-order sedimentary sequence. This sedimentary sequence in each of the four studied wells has a varying thickness ranging from about 106 to 141 meters. Considering the fact that the age of this sequence is Coniacian-Santonian, therefore, it is regarded as a third-order sedimentary sequence. Two system tracts HST and TST and also MFS in this formation were recognized.

(4) Based on the investigations of reservoir quality of the Ilam Formation in one of the studied wells, 6 flow units were identified. Flow unit 5 has the best reservoir quality and flow unit 6 is the worst.

## Acknowledgments

This study is an industry project. The authors are grateful to the National Iranian Oil Company for support and data preparation.